\documentclass[twocolumn,aps,prb,amsmath,amssymb,preprintnumbers,superscriptaddress,longbibliography]{revtex4-2}

\usepackage[utf8]{inputenc}
\usepackage{physics}
\usepackage[normalem]{ulem}
\usepackage{braket}
\usepackage{bbold}
\usepackage{amsmath, amsfonts, amssymb}
\usepackage{graphicx}
\usepackage{float}
\usepackage{lipsum}
\usepackage[colorlinks=true,linkcolor=blue,citecolor=blue,allcolors=blue]{hyperref}
\usepackage[caption=false]{subfig}
\usepackage{mathtools}
\usepackage{tikz}
\usepackage{natbib}

\begin{document}
	\newcommand{\titleinfo}{Simulating Quantum Circuits with Tree Tensor Networks using Density-Matrix Renormalization Group Algorithm}
	\title{\titleinfo}

        \author{Aditya Dubey}
        \email{f20200627@goa.bits-pilani.ac.in}
        \affiliation{Department of Physics, BITS-Pilani, K K Birla Goa Campus, Zuarinagar, Goa 403726, India}
        \affiliation{Zentrum f{\"u}r Optische Quantentechnologien, Universit{\"a}t Hamburg, Luruper Chaussee 149, 22761 Hamburg, Germany}
 
        \author{Zeki Zeybek}
	\email{zeki.zeybek@uni-hamburg.de}
	\affiliation{Zentrum f{\"u}r Optische Quantentechnologien, Universit{\"a}t Hamburg, Luruper Chaussee 149, 22761 Hamburg, Germany}
        \affiliation{The Hamburg Centre for Ultrafast Imaging, Universit{\"a}t Hamburg, Luruper Chaussee 149, 22761 Hamburg, Germany}
	
        \author{Peter Schmelcher}
        \email{peter.schmelcher@uni-hamburg.de}
\affiliation{Zentrum f{\"u}r Optische Quantentechnologien, Universit{\"a}t Hamburg, Luruper Chaussee 149, 22761 Hamburg, Germany}	
 \affiliation{The Hamburg Centre for Ultrafast Imaging, Universit{\"a}t Hamburg, Luruper Chaussee 149, 22761 Hamburg, Germany}
	
	\begin{abstract}
    Quantum computing offers the potential for computational abilities that can go beyond classical machines. However, they are still limited by several challenges such as noise, decoherence, and gate errors. As a result, efficient classical simulation of quantum circuits is vital not only for validating and benchmarking quantum hardware but also for gaining deeper insights into the behavior of quantum algorithms. A promising framework for classical simulation is provided by tensor networks. Recently, the Density-Matrix Renormalization Group (DMRG) algorithm was developed for simulating quantum circuits using matrix product states (MPS). Although MPS is efficient for representing quantum states with one-dimensional correlation structures, the fixed linear geometry restricts the expressive power of the MPS. In this work, we extend the DMRG algorithm for simulating quantum circuits to tree tensor networks (TTNs). The framework employs a variational compression scheme that optimizes the TTN to approximate the evolved quantum state. To benchmark the method, we simulate random circuits and the quantum approximate optimization algorithm (QAOA) with various two-qubit gate connectivities. For the random circuits, we devise tree-like gate layouts that are suitable for TTN and show that TTN requires less memory than MPS for the simulations. For the QAOA circuits, a naive TTN construction that exploits graph structure significantly improves the simulation fidelities. Our findings show that the DMRG algorithm with TTNs provides a promising framework for simulating quantum circuits, particularly when gate connectivities exhibit clustering or a hierarchical structure.
	\end{abstract}
	\maketitle

\section{Introduction} 

Quantum computing promises revolutionary computational capabilities, offering the potential to solve problems that are intractable for classical machines \cite{Feynman1982,Shor,Chem_QC}. However, despite the allure of exponentially faster algorithms, quantum computers today are still hampered by several challenges \cite{Preskill2018quantumcomputingin,Quantum_House_of_Cards}.  Current devices are limited by noise, decoherence, and gate errors, all of which constrain their reliability. As a result, efficient classical simulation of quantum systems is vital for validating and benchmarking quantum hardware and gaining deeper insights into the behavior of quantum algorithms \cite{TEBD_Miles,FFT_Miles,Grover_Miles}. 

A promising framework for classical simulation is provided by tensor networks (TNs) \cite{SCHOLLWOCK_TN_bible,ORUS_TN,TN_review_Carmen,DMRG_review}. They offer a way to manage the exponential complexity of quantum states by exploiting the structure inherent in many-body entanglement, rendering them ideal for the representation and manipulation of quantum states. Consequently, TN algorithms have become cornerstones in the numerical study of strongly correlated quantum systems, addressing problems ranging from ground state search \cite{DMRG_White,DMRG_Schollwock} to real-time dynamics \cite{Vidal_Slightly_entangled, tdvp_mps, TimeEvolution_Schollwock}. 

Tensor networks have also become a versatile tool in quantum information theory and quantum computing \cite{TN_QC_Rev}. For example, they provide a framework for loading classical data into a quantum circuit \cite{classicalData_TN,GenModel_MPS,ML_TN_Miles,MPS_prep_original_Cirac,state_prep_riemann_opt,Const_depth_MPS}. TN techniques also provide valuable insights into the structure and performance of quantum algorithms \cite{Grover_Miles, TTN_Shor, FFT_Miles}. Efficient classical simulation of quantum circuits is equally vital for the validation and benchmarking of quantum hardware. TN methods have proven particularly effective in this context by refuting quantum supremacy claims with classical simulation results that rival or surpass proposed quantum advantages \cite{QC_simulation_Batch_TN,TEBD_Miles,DMRG_Miles,Miles_kicked_ising,Gaussian_boson_sampling_TN}.

A wide variety of tensor network architectures have been employed for simulating quantum circuits \cite{TEBD_Miles,DMRG_Miles,TTN_Mendl,QC_MERA,PEPS_random_circuits}. In particular, the DMRG algorithm using matrix product states was generalized to simulate quantum circuits with a finite fidelity \cite{DMRG_Miles}. The algorithm essentially takes the resulting state after the application of the gates and compresses it into an MPS. Although MPS is efficient for compressing quantum states that admit one-dimensional correlation structure, the fixed chain-like geometry restricts the expressive power of the MPS. Therefore, an appropriate kind of tensor network should be considered for the DMRG algorithm to represent states exhibiting multi-scale correlations. 

Tree tensor networks have emerged as a powerful tool in both many-body physics and quantum chemistry, with applications spanning ground state calculations \cite{TTN_MBP1, TTN_MBP2,TTN_MBP3, TTN_MBP4,TTN_MBP5,TTN_MBP6,TTN_GS_Rizzi}, real-time dynamics \cite{TTN_Vidal,TTN_TEBD1,TTN_TEBD2,TTN_dynamics1,TTN_dynamics2,ML_Spin_Dynamics,TTN_Dynamics_Rizzi,Dynamics_TTN_Rizzi}, and the study of complex molecular systems \cite{TTN_Qchem1,TTN_Qchem2,TTN_Qchem3,TTN_Qchem4}. However, TTNs have not been explored extensively in the context of quantum circuit simulation. Their structural flexibility allows for adapting the ansatz to match the underlying connectivity or correlation patterns of the quantum system. {A direct application of gates on a TTN leads to a complex tensor network structure, which would then require transformation back into the TTN shape. However, naive transformation of a tensor network can result in an exponentially large memory requirement, scaling with the number of qubits. Ref.~\cite{TTN_Mendl} addresses this challenge by introducing the concept of \textit{threading} entanglement through the tree. This method is equivalent to approximating the true output quantum state using a truncation method based on singular value decomposition (SVD). The truncation is based on discarding small singular values which can introduce a notable error with each gate. These errors compound with subsequent gate applications, potentially leading to a significant loss of fidelity.}

{In this work, we extend the Density-Matrix Renormalization Group (DMRG) algorithm, originally developed for simulating quantum circuits with MPS \cite{DMRG_Miles}, to TTNs. The central idea of this method is a variational compression scheme that optimizes the TTN to approximate the evolved quantum state. It iteratively optimizes each tensor to find the best possible TTN representation of the evolved quantum state for a given bond dimension. This variational approach is superior to the SVD-based compression method as it minimizes the distance to the target state and avoids the compounding errors associated with successive SVD truncations \cite{SCHOLLWOCK_TN_bible,DMRG_Miles}. Furthermore, the framework preserves the shape of the TTN upon the application of gates and enables dynamical change of the TTN structure mid-simulation.} The key implication of our work is that {the DMRG algorithm with} TTNs provides a promising framework for simulating quantum circuits, especially when gate connectivities exhibit some form of clustering or hierarchical structure. Therefore, the flexibility of TTNs allows them to adapt to a wide range of quantum circuit architectures. We focus on random circuits and QAOA circuits with various connectivities, including a special class of circuits referred to as \textit{tree-like} circuits. For tree-like random circuits, TTNs require less memory than MPS for comparable accuracy due to their ability to better capture long-range correlations. In QAOA simulations for MaxCut on 3-regular and tree-like connected graphs, a naive TTN construction exploiting graph structure significantly improves fidelity compared to a blind approach. We achieve near-perfect fidelity for structured bridged 3-regular and tree-like connected graphs despite having small bond dimensions.


The paper is organized as follows. In Sec.~\ref{Sec: TTN}, we introduce the TTN ansatz and outline its key features. Section~\ref{Sec: Compression} presents the DMRG algorithm as applied to the simulation of quantum circuits. In Sec.~\ref{Sec: Results}, we report our results for the simulation of both random and QAOA circuits across various qubit connectivity configurations. Finally, in Sec.~\ref{Sec: Conclusion} we provide our conclusions and outlook. {Appendix \ref{Appendix: SVD compare} presents a comparison between the DMRG algorithm and the SVD-based truncation scheme for simulating quantum circuits, and Appendix \ref{Appendix: Ansatz compare} provides a comparison between MPS and TTN within the DMRG framework.}

\section{Theory}
In the following, the tree tensor network (TTN) ansatz is introduced and we explain the DMRG algorithm for simulating quantum circuits. 

\begin{figure}[!t]
    \centering
    \includegraphics[width=1\columnwidth]{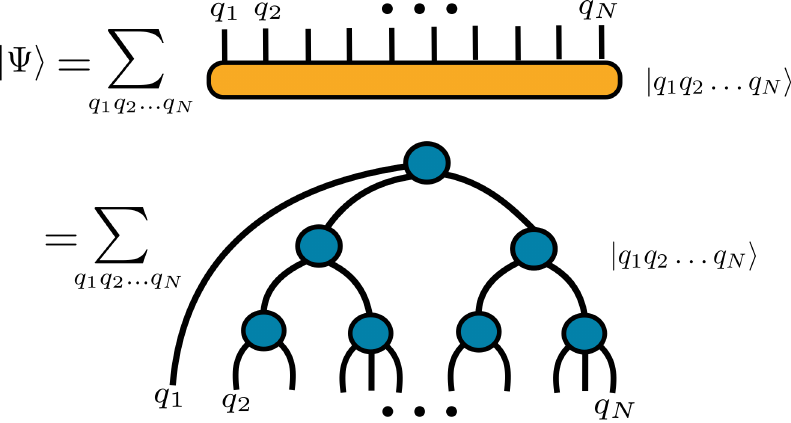}
   \caption{Illustration of the TTN ansatz is given. TTN represents the coefficient tensor (orange bar) as a  
   hierarchical network of tensors with a tree structure. Fully contracting the network down to the leaf nodes with a specific set of physical indices yields the coefficient of the corresponding basis state.} 
   \label{Fig: TTN ansatz}
\end{figure}

\begin{figure*}
\centering
\includegraphics[width=\textwidth]{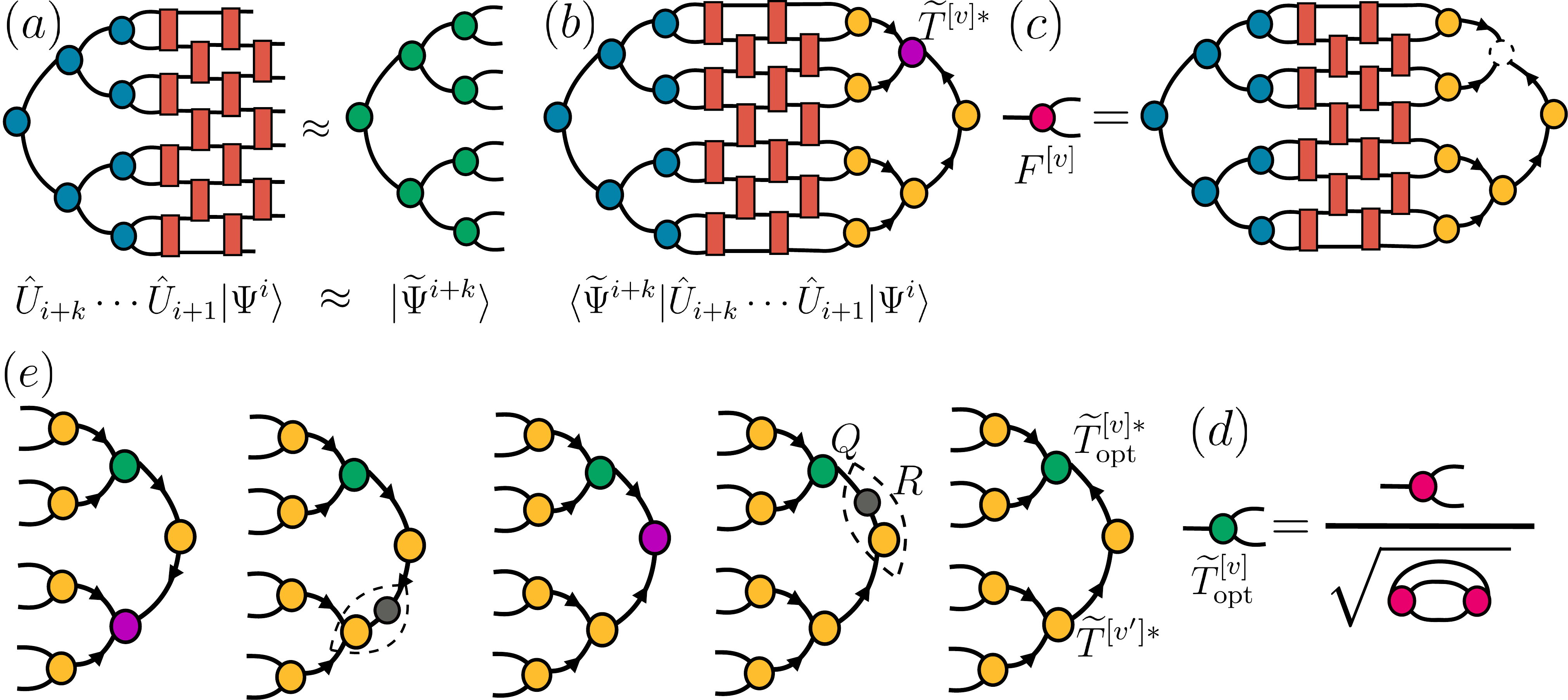}
   \caption{DMRG algorithm for simulating quantum circuits is illustrated. (a) A schematic of the compression step: \(k\) gates are applied to the initial state \(\ket{\Psi^i}\), followed by an approximation of the resulting state with a tree tensor network \(\ket{\widetilde\Psi^{i+k}}\). 
(b) Tensor network diagram of the scalar product being maximized; the purple node indicates the tensor currently being optimized, which serves as the orthogonality center. 
(c--d) The optimal tensor \(\widetilde{T}^{[v]}_{\text{opt}}\) is computed from the environment tensor \(F^{[v]}\), normalized by \(\sqrt{\text{Tr}F^{[v]} F^{[v]*}}\). 
(e) The orthogonality center is shifted to the next tensor to be optimized, \(\widetilde{T}^{[v^\prime]}\), via a sequence of QR decompositions.
}
\label{Fig: DMRG Aglorithm}
\end{figure*}

\subsection{Tree tensor network}
\label{Sec: TTN}
An arbitrary quantum state $\ket{\Psi}$ describing a system of $N$ qubits is expressed as 
\begin{equation}
    \label{eqn:state_vec}
    \ket{\Psi} = \sum_{q_1q_2\dots q_N} \Psi_{q_1q_2\dots q_N}\ket{q_1q_2\dots q_N},
\end{equation}
where $\Psi_{q_1q_2\dots q_N}$ is the coefficient tensor and $\ket{q_1q_2\dots q_N}$ denotes the basis states. The coefficient tensor requires $2^{N}$ parameters to be specified exhibiting exponential scaling with the system size. 
A TTN decomposes this large tensor into a product of smaller tensors with a hierarchical tree structure where the individual tensors are located at the nodes of a connected acyclic graph and the edges between nodes indicate how these tensors are contracted as shown in Fig.~\ref{Fig: TTN ansatz}. The TTN ansatz is given as

\begin{equation}
\label{eqn: TTN_ansatz}
    \Psi_{q_1q_2\dots q_N} =\sum_{\{\chi\}}T_{\{q\}_1,\{\chi\}_1}^{[1]} \cdots T_{\{q\}_v,\{\chi\}_v}^{[v]} \cdots T_{\{q\}_l,\{\chi\}_l}^{[l]},
\end{equation}
where the summation is over all the virtual indices collectively denoted by $\{\chi\}$. {$T_{\{q\}_v,\{\chi\}_v}^{[v]}$} is the tensor associated with the node $v$ on the tree network, $\{\chi\}_v$ is the set of virtual indices that are connected to the node and  $\{q\}_v$ represents the set of physical indices that are connected to the node. Note that $\{q\}_v$ can be an empty set, whereas $\{\chi\}_v$ is always non-empty as every node in the TTN is connected to the rest of the network through its virtual bonds. {In this study, we employ layer-wise regular tree structures for the TTN ansatz. Specifically, all non-leaf nodes at a given layer have the same number of children, though this number can vary across layers. This provides more flexibility than a strictly regular tree, where every non-leaf node must have the same number of children.} The total number of indices (i.e., elements of $\{q\}_v \cup {\{\chi\}}_v$) for each tensor in the TTNs is at most five { with at most four virtual indices. Conventionally, TTNs are built using tensors with at most three virtual legs to control scaling with the bond dimension \cite{TN_Anthology_Silvi}. We relax this restriction for two main reasons. First, such a constraint severely limits the flexibility in choosing tree topologies for arbitrary qubit counts. For instance, creating a layer-wise regular TTN for a 27-qubit system becomes impractical under this condition. Second, when the qubit connectivity exhibits a hierarchical clustering, it is beneficial to capture that structure at higher levels of the tree. This may require tensors with more than three virtual legs to efficiently group strongly correlated qubits. Such clustering helps keep bond dimensions manageable, which in turn can mitigate the additional scaling overhead introduced by allowing more virtual legs.}

TTNs provide a loop-free tensor network representation that allows full exploitation of tensor network gauge freedom.{ 
This gauge freedom allows the TTN to be brought into the so-called canonical form \cite{Unconstrained_TTN, TN_Anthology_Silvi}. In this form, one tensor is designated as the orthogonality center, and all other tensors are isometric with respect to the direction pointing toward it.} Smart gauging can significantly reduce the number of required tensor contractions and thereby improve computational efficiency. {For example, the norm of the TTN reduces to a single contraction of the orthogonality center with its conjugate.} This feature is crucial for the DMRG algorithm used in this study as it greatly simplifies the compression step, as discussed in Sec.~\ref{Sec: Compression}. In contrast to MPS, the distance between physical indices in a TTN scales logarithmically with the system size as $\mathcal{O}(\log N)$, whereas in MPS, it scales linearly $\mathcal{O}(N)$. Since connected correlation functions typically decay exponentially with the path length in a tensor network, the shorter paths in TTNs allow for a more faithful representation of long-range correlations compared to MPS \cite{TTN_Mendl,TTN_MBP1}. This property makes TTNs more efficient at representing systems with long-range interactions or quantum circuits with non-local gates. In this work, we leverage this flexibility to simulate a wide class of circuits with a carefully constructed TTN.

\subsection{Quantum circuit simulation}
\label{Sec: Compression}

In the DMRG algorithm for simulating quantum circuits, the main idea is to approximate the evolving quantum state using a TTN ansatz as illustrated in Fig.~\ref{Fig: DMRG Aglorithm}(a). {The approximation is performed through a variational compression scheme by iteratively optimizing individual tensors to minimize the distance to the target state. This approach is fundamentally superior to the SVD-based method as it avoids the accumulation of truncation errors \cite{SCHOLLWOCK_TN_bible,DMRG_Miles}. We demonstrate this explicitly in Appendix \ref{Appendix: SVD compare} by comparing our approach with the one given in \cite{TTN_Mendl}.}

To formalize our algorithm, let us consider a TTN at step \( i \) denoted by \( \ket{\Psi^i} \), which ideally approximates the exact quantum state after the application of \( i \) gates. When a sequence of \( k \) additional gates, \( \hat{U}_{i+k} \cdots \hat{U}_{i+1} \), is applied, the updated state is given as
\begin{equation} \ket{\Psi^{i+k}} = \hat{U}_{i+k} \cdots \hat{U}_{i+1} \ket{\Psi^i}. 
\end{equation} 
{The goal is to approximate this evolved state with a new TTN \( \ket{\widetilde\Psi^{i+k}} \), whose structure can be chosen arbitrarily and is independent of the previous TTN \( \ket{\Psi^{i}} \). For all the simulations in this work, we keep the structure of TTN fixed. The approximation is obtained by minimizing the distance 
\begin{equation}
\delta = \left\| \ket{\Psi^{i+k}} - \ket{\widetilde\Psi^{i+k}} \right\|^2, 
\end{equation} 
 through iterative optimization of each tensor \( \widetilde{T}^{[v]} \) in \( \ket{\widetilde\Psi^{i+k}} \) individually while keeping the others fixed.}
Each such optimization step reduces to solving a linear system of equations, whose solution yields the optimal tensor \( \widetilde{T}^{[v]}_{\text{opt}} \). 
To simplify this optimization, it is advantageous to put \( \ket{\widetilde\Psi^{i+k}} \) into a canonical form with respect to the target tensor \( \widetilde{T}^{[v]} \), as illustrated in Fig.~\ref{Fig: DMRG Aglorithm}(b). 
In this form, all other tensors in the network are isometric in the direction of \( \widetilde{T}^{[v]} \), which serves as the orthogonality center. 
This ensures that the norm of the TTN is simply given by \begin{equation} \braket{\widetilde\Psi^{i+k} | \widetilde\Psi^{i+k}} = \text{Tr}\widetilde{T}^{[v]} \widetilde{T}^{[v]*},
\end{equation}
where the Tr denotes the full contraction of the network with summation over all indices. The optimal target tensor (up to normalization) is then obtained from the environment tensor \( F^{[v]} \), which is computed by contracting the overlap network \( \braket{\widetilde\Psi^{i+k} | \Psi^{i+k}} \) without the tensor \( \widetilde{T}^{[v]*} \) \cite{SCHOLLWOCK_TN_bible}, as shown in Fig~\ref{Fig: DMRG Aglorithm}(c--d). The resulting expression for the optimal target tensor is given as
\begin{equation}
    \widetilde{T}^{[v]}_{\text{opt}} = \frac{F^{[v]}}{\sqrt{\text{Tr}F^{[v]} F^{[v]*}}}.
\end{equation}
To monitor convergence during the optimization of each tensor, the distance can be monitored $ \delta_v = 1 - f_v,$
where the partial fidelity \( f_v \) is defined as  $f_v = \text{Tr}F^{[v]} F^{[v]*}. $
This fidelity also serves as a measure of how well the updated TTN approximates the evolved state,
\begin{equation} f_v = \left\| \braket{\widetilde\Psi^{i+k} | \hat{U}_{i+k} \cdots \hat{U}_{i+1} | \Psi^i} \right\|^2, \end{equation}
enabling us to quantitatively assess the progress of the optimization process during individual tensor updates or across multiple sweeps. In this iterative process, the partial fidelities increase monotonically as we continuously improve the TTN by updating the tensors. 

This procedure allows the simulation of an entire quantum circuit by decomposing it into a sequence of compression steps. The final value of the partial fidelity $f^{[\kappa]}$ obtained at each compression step $\kappa$ can be used to define the quantity
\begin{equation} \mathcal{\tilde{F}} = \prod_{\kappa} f^{[\kappa]}. \end{equation}
As demonstrated in Refs.~\cite{TEBD_Miles, DMRG_Miles}, $\mathcal{\tilde{F}}$ serves as an accurate approximation of the exact fidelity between the exact and compressed states. Since we employ the same algorithm as in Ref.~\cite{DMRG_Miles}, this approximation remains valid in our case. Based on this, we define the error rate per two-qubit gate in our simulation as \begin{equation}
\label{Eq: error}
 \epsilon = 1 - \mathcal{\tilde{F}}^{1/N_{2g}},
    \end{equation}
where $N_{2g}$ denotes the total number of two-qubit gates. This quantity is commonly employed as a benchmark for evaluating the performance of NISQ devices \cite{Google_supremacy,QC_trapped_ion,QC_advantage}.

\section{Results}
\label{Sec: Results}
In the following, the results of simulations for random circuits and QAOA circuits across various qubit connectivity configurations are presented. For all the simulations, the initial state is set to $\ket{00\cdots00}$.

\subsection{Random circuits}
\label{sec: Results random}
Although random circuits are not typically designed to solve practical computational problems, they play a crucial role as benchmarks for demonstrating quantum advantage on near-term quantum devices \cite{Random_Circuit_quantum_complexity,QuantumAdvantage_Quantum_simulation,Google_supremacy}. Their significance lies in the high degree of entanglement they generate, which poses a major challenge for classical simulation methods such as tensor networks that rely on low-entanglement structures. In these simulations, we focus on the error per gate $\epsilon$ {as defined in Eq.~\ref{Eq: error}}. A very recent work by Quantinuum demonstrated random circuit implementations on arbitrary geometries \cite{Random_circ_Quantinuum}, which they benchmarked using the DMRG algorithm with MPS using error per gate  $\epsilon$ as the metric. However, the linear structure of MPS limits its expressive power for such circuits. In this context, TTN may offer a more suitable framework for benchmarking these devices.

In the following,  we test the method on random circuits where each two-qubit gate making up the circuit is generated through a randomized two-step procedure. Initially, the real and imaginary components of a \( 4 \times 4 \) matrix are independently drawn from a normal distribution over the interval $[0,1]$. Next, the sampled matrix undergoes QR decomposition, yielding \( A = QR \), where \( Q \) is a unitary matrix and \( R \) is an upper triangular matrix. The resulting unitary matrix \( Q \) is then used as the two-qubit gate. 

\begin{figure}[!t]
    \centering
    \includegraphics[width=1\columnwidth]{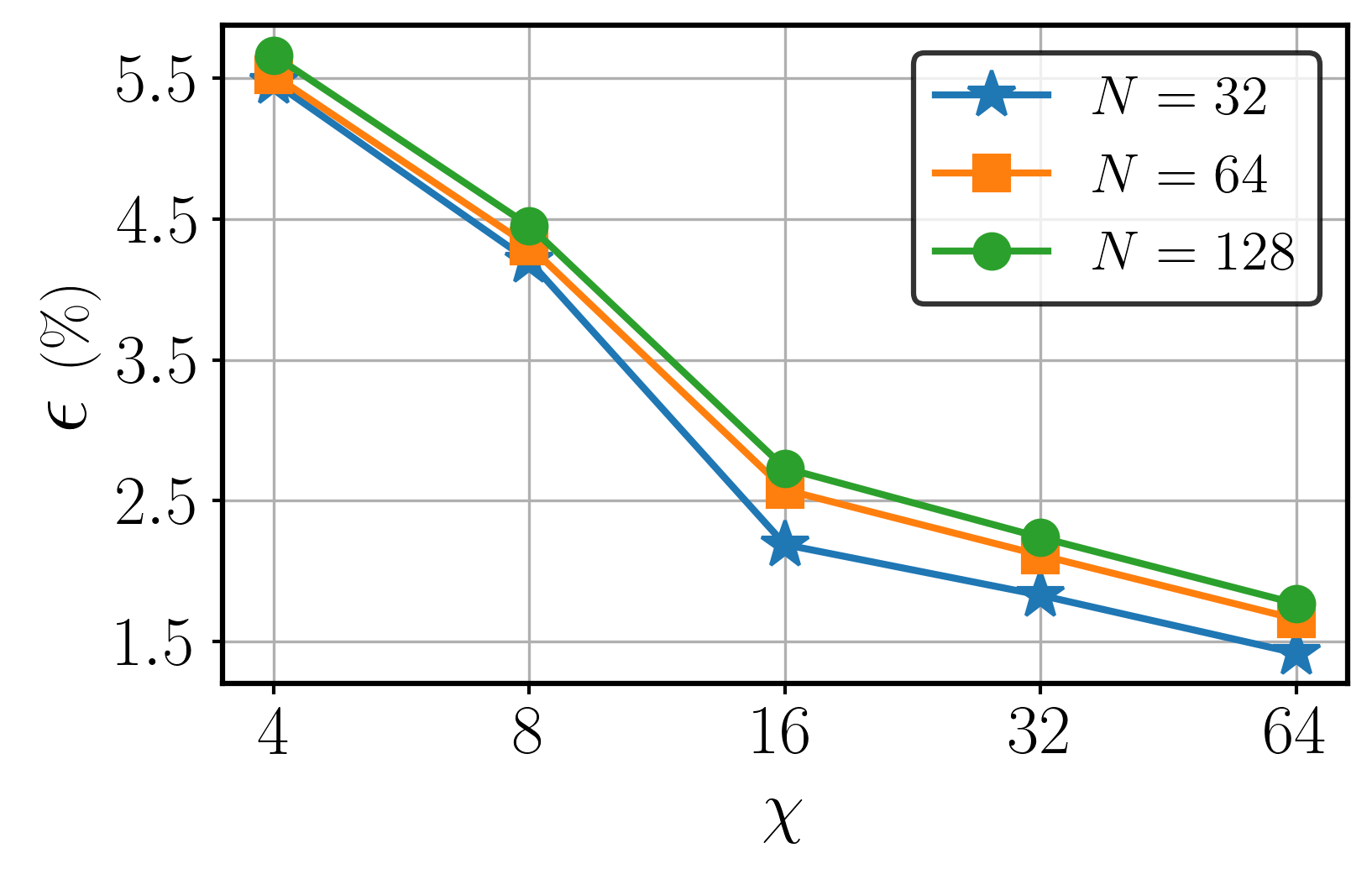}
    \caption{Error per gate $\epsilon$ as a function of bond dimension $\chi$ obtained from the simulation of random brick-wall circuits. The results are shown for $N_{2g}=310,630,$ and $1270$ two-qubit gates, corresponding to systems of $N=32,64,$ and $128$ qubits, respectively. } 
    \label{Fig: Error NN}
\end{figure}

\subsubsection{Nearest-neighbor connectivity}

We first consider quantum circuits with nearest-neighbor connectivity among two-qubit gates. In these circuits, the gates are placed in a brick-wall layout with each two-qubit gate acting on adjacent qubits arranged in a chain-like geometry. In terms of two-qubit connectivity, this can be considered as a linear graph where the vertices denote the qubits and the edges between them are the two-qubit gates. At a given circuit depth $D$, the total number of two-qubit gates is given by $N_{2g} = N - 1$, which is the total number of edges on the two-qubit gate connectivity graph. To simulate these circuits, we employ a TTN with a binary structure with leaf nodes accommodating four qubits instead of two. In this TTN ansatz, the maximum bond dimension $\chi$ achievable is given by $\chi_{\max} = 2^{N/2}.$  

\begin{figure}[!t]
    \centering
    \includegraphics[width=1\columnwidth]{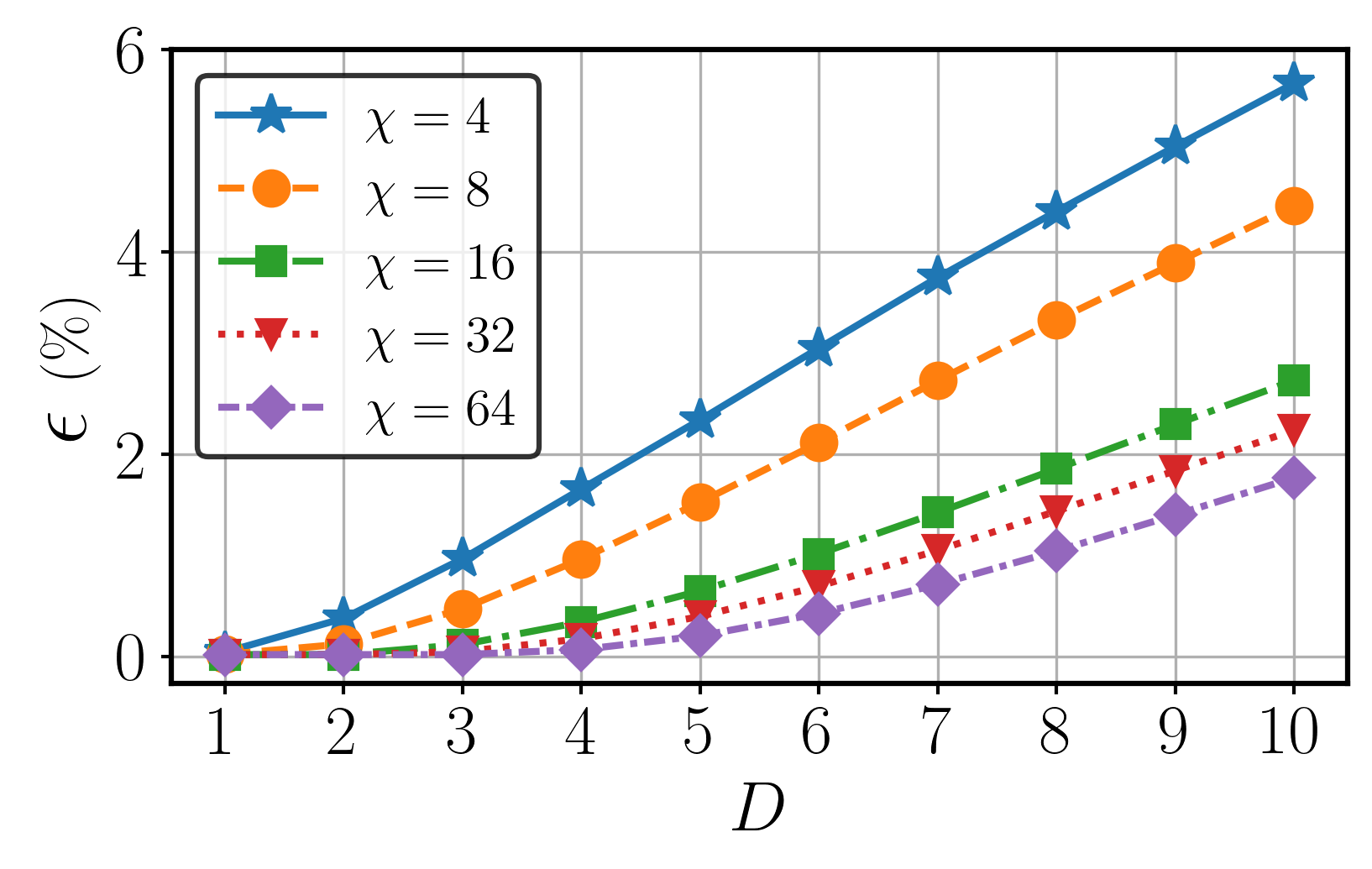}
    \caption{Error per gate $\epsilon$ as a function of the circuit depth $D$ obtained from the simulation of random brick-wall circuits with $128$ qubits. Each depth contains $127$ two-qubit gates, resulting in a total of $N_{2g}=1270$. Different curves correspond to different values of the bond dimension $\chi$ ranging from 4 to 64.} 
    \label{Fig: Error NN 128}
\end{figure}

 Fig.~\ref{Fig: Error NN} shows the error per gate $\epsilon$ as a function of the bond dimension $\chi$ for random circuits with a depth of $D = 10$ and $N = 32, 64,$ and $128$ number of qubits, where the total number of two-qubit gates are given by $N_{2g}=310,630,$ and $1270$, respectively. At a moderate bond dimension of $\chi=64$, which is much smaller than $\chi_{\max}$, errors per gate of $\epsilon \approx 1.5\%$ is achieved for all the simulation instances. For the largest system size with $N=128$, the behavior of the error per gate $\epsilon$ as a function of the circuit depth $D$ for various bond dimensions $\chi$ is shown in Fig.~\ref{Fig: Error NN 128}. As expected, increasing the bond dimension leads to a reduction in error growth rates with respect to the circuit depth. It is also worth mentioning that matrix product states (MPS) would likely be a more suitable choice than TTNs for simulating these circuits. The linear geometry of the MPS ansatz does not impose any bias as to what group of qubits would be more entangled by the nearest-neighbor two-qubit connectivity. Instead, it provides a more natural tensor network representation for one-dimensional chains with short-ranged interactions. With a given TTN structure, a given hierarchical grouping of the qubits may not be optimal for representing the entanglement generated by random circuits with nearest-neighbor connectivity.

\begin{figure}[!h]
\centering
\includegraphics[width=1\columnwidth]{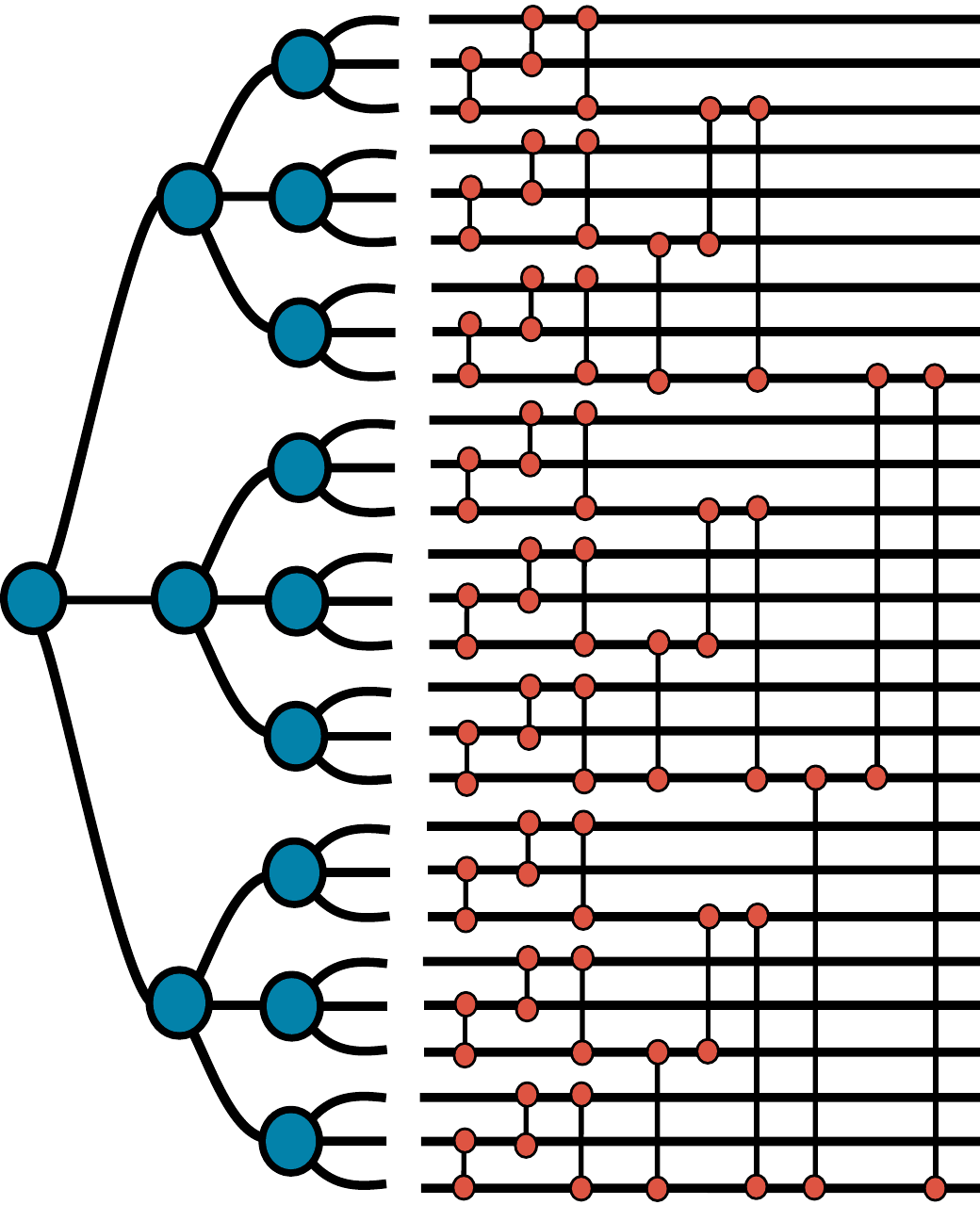}
\caption{Single depth of a tree-like circuit resulting from a ternary tree tensor network of 27 qubits. In this circuit, qubits associated with the nodes of the bottom layer of the TTN are fully connected, forming nine clusters of three qubits each. These clusters are then connected to clusters associated with the same parent node in the second layer of the TTN via single two-qubit gates, specifically acting on the first qubit of each cluster. This results in larger clusters comprising nine qubits. At the highest level of the hierarchy, these nine-qubit clusters are further connected using a single two-qubit gate between the first qubit of each cluster, thereby completing one depth of the tree-like circuit.
} 
\label{Fig: Tree-like circuit}
\end{figure}

\subsubsection{Tree-like circuit connectivity}
 In this section, we consider a circuit with a built-in structure that admits a hierarchical clustering of the qubits. Specifically, we investigate a circuit layout that mirrors the TTN structure that represents the quantum state. We refer to such circuits as \textit{tree-like circuits} \cite{TTN_Mendl} since their gate connectivity forms a graph exhibiting a hierarchical tree-like clustering as shown in Fig.~\ref{Fig: Tree-like circuit} for a ternary tree structure with $N=27$ qubits. In these circuits, qubits associated with the nodes of the bottom layer of the TTN are all-to-all connected, forming clusters. Each of these clusters represents a region of strong interaction, where two-qubit gates are applied among all qubits within the cluster. These initial clusters are then merged into larger clusters at the next hierarchical level. In our construction, the inter-cluster connectivity at every level is provided through a single two-qubit gate. A single two-qubit gate acting on the first qubit of each cluster establishes the couplings between the larger clusters. This process is repeated iteratively, where each layer aggregates smaller clusters into larger ones, forming a hierarchy of clusters that reflects the TTN structure. 

\begin{figure}[!t]
    \centering
    \includegraphics[width=1\columnwidth]{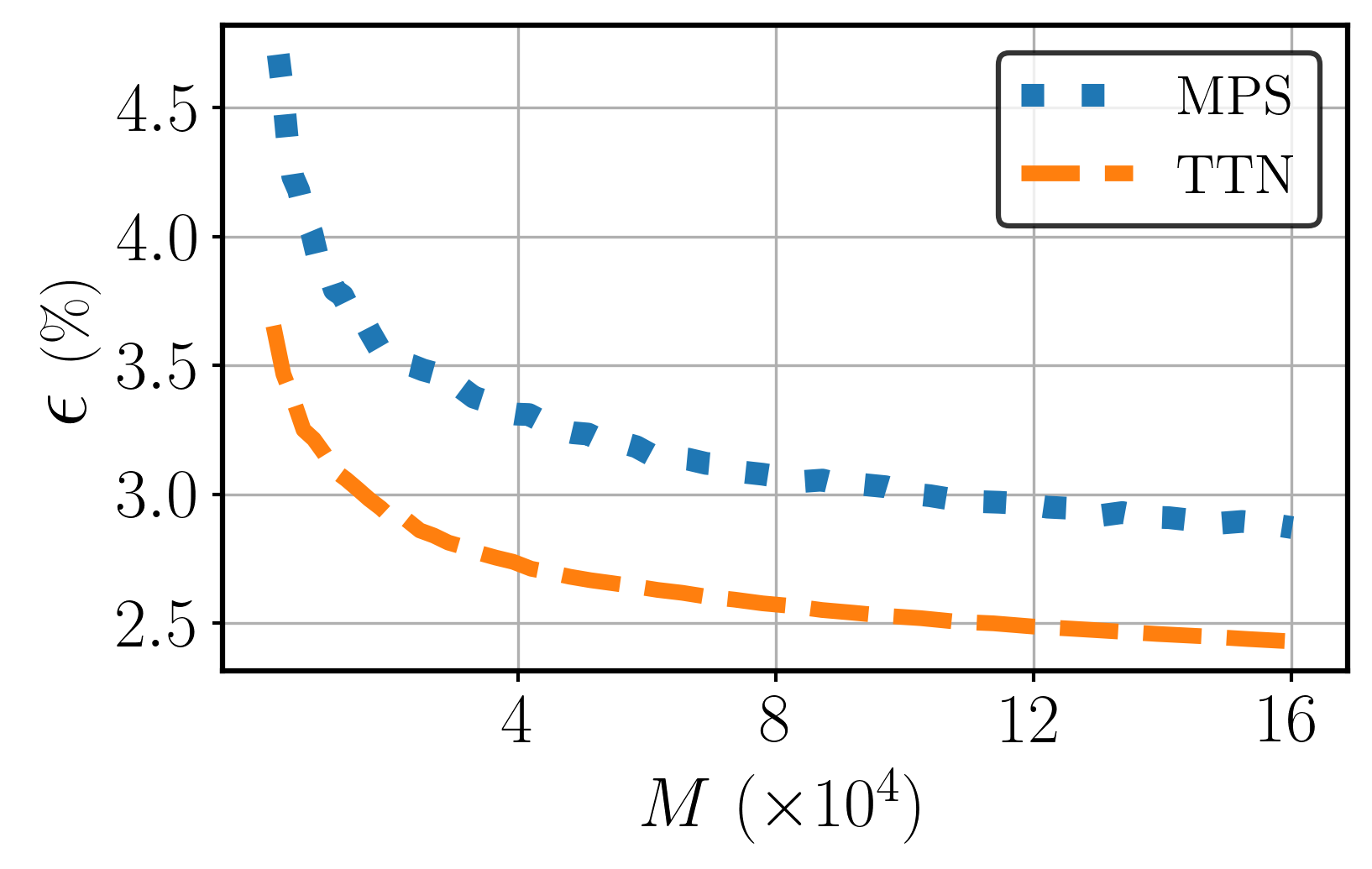}
    \caption{Comparison of error rates $\epsilon$ as a function of memory footprint $M$ obtained from the simulation of random tree-like circuits using MPS and TTN. The results are for a system of $N=27$ qubits with $N_{2g}=390$ two-qubit gate.} 
    \label{Fig: ErrorVsMem}
\end{figure}

In Fig.~\ref{Fig: ErrorVsMem}, we analyze the error per gate versus memory footprint $M$ for the simulation of these circuits. The memory footprint $M$ is defined as the total number of tensor elements $M = \sum_{v}\prod_{i} \text{dim}_i(T^{[v]})$. Here, $\text{dim}_i(T^{[v]})$ refers to the dimension of index $i$ of the tensor associated with the node $v$. This analysis focuses on simulating random tree-like circuits with $N_{2g}=390$ for $N=27$ qubits using MPS and TTN. These circuits pose a significant challenge for MPS, as achieving comparable gate errors to TTN requires substantially more memory. This difficulty stems from the inherent limitation of MPS in efficiently capturing non-local correlations induced by long-range gates due to its strictly linear topology. Specifically, when two qubits that are far apart in the MPS chain are entangled by a gate then the intermediate tensors must carry this non-local information. To faithfully encode these correlations, the bond dimension must grow across the path connecting the two qubits, resulting in a greater memory footprint. In contrast, TTN leverages the underlying connectivity of the
circuit to construct a more expressive ansatz that allows it to capture correlations more effectively while requiring less memory. The results highlight that when gate connectivity can be exploited, TTN can be a more efficient choice than MPS in simulating such quantum circuits.

\begin{figure}[!t]
    \centering
    \includegraphics[width=1\columnwidth]{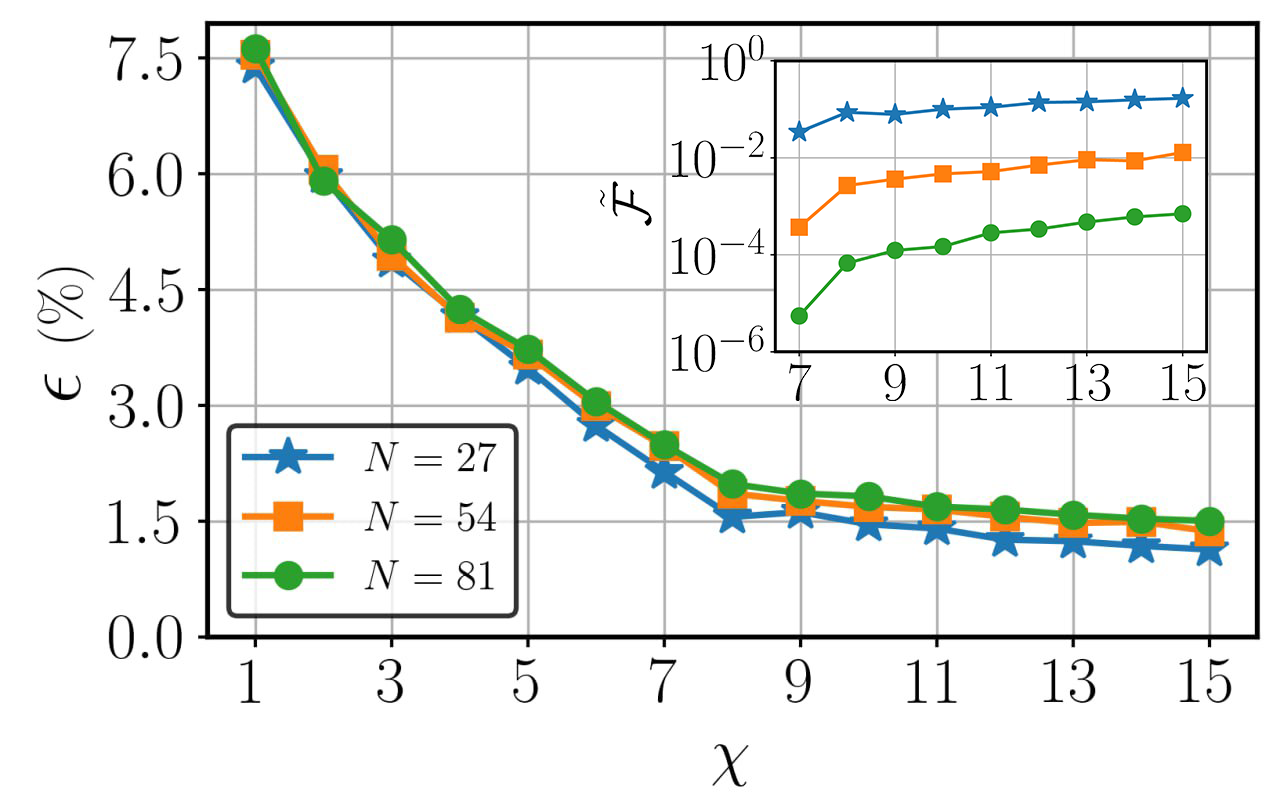}
    \caption{Error per gate $\epsilon$ as a function of bond dimension $\chi$ obtained from the simulation of random tree-like circuits. The results are shown for $N_{2g}=156,326,$ and $480$ two-qubit gates, corresponding to systems of $N=27,54,$ and $81$ qubits. The inset shows fidelity $\mathcal{\tilde{F}}$ as a function of bond dimension $\chi$ obtained from the same simulations.} 
    \label{Fig: Error v D for TL}
\end{figure}

In Fig.~\ref{Fig: Error v D for TL}, the error per gate is shown as a function of the bond dimension $\chi$ for the simulation of random tree-like circuits with system sizes of $N = 27$, $54$, and $81$ qubits. The corresponding total number of two-qubit gates is $N_{2g} = 156$, $316$, and $480$, respectively. Notably, an error of $\epsilon \approx 1.5\%$ is achieved even for a small bond dimension of $\chi = 15$ for all the simulations. The inset displays the fidelities $\mathcal{\tilde{F}}$ from the same simulations. This suggests that having similar error values $\epsilon$ across circuits with different qubit counts does not ensure their fidelities will be comparable. Therefore, as the qubit count increases, low fidelities are likely at such small bond dimensions due to the high entanglement generated in these circuits.

\begin{figure}[!t]
\centering
\includegraphics[width=1\columnwidth]{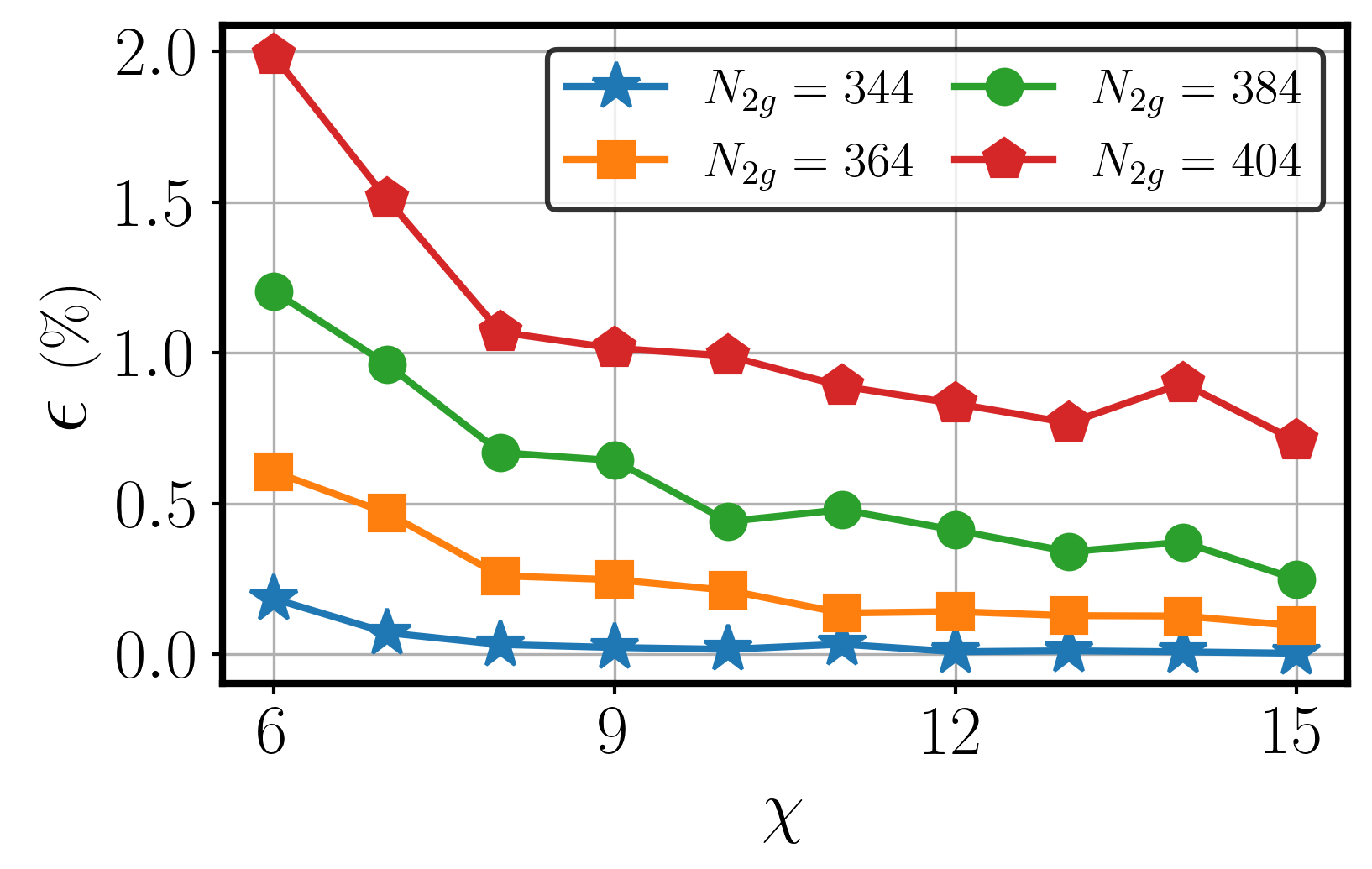}
\caption{Error per gate $\epsilon$ as a function of bond dimension $\chi$ obtained from the simulation of random tree-like circuits for a system of $N=81$ qubits with varying number $N_{2g}$ of two qubit gates. } 
\label{Fig: Sample Random TL}
\end{figure}

To understand how the error per gate $\epsilon$ depends on the degree of inter-cluster connectivity, tree-like circuits with different numbers of inter-cluster gate counts are simulated. Specifically, we focus on a tree-like circuit based on a ternary tree structure with $N = 81$ qubits. We randomly select a specific number of two-qubit gates to connect different clusters. This selection is made uniformly from all possible connections between clusters with repetitions allowed. In all cases, the total number of intra-cluster two-qubit gates is kept all-to-all, and only the inter-cluster gates are varied. In Fig.~\ref{Fig: Sample Random TL}, the error per gate $\epsilon$ as a function of $\chi$ for different numbers of two-qubit gates is shown. Despite having a random circuit, TTN allows us to achieve very low error per gate of $\epsilon\approx 0.003 \%$ by appropriately grouping strongly interacting qubits in one cluster as shown by the results for $N_{2g}=344$ (Fig.~\ref{Fig: Sample Random TL}). As the number of inter-cluster gates increases, the error per gate increases, as is expected due to the corresponding increase in the entanglement of inter-cluster qubits. It should also be noted that the increase in error per gate cannot be solely attributed to the number of inter-cluster gates. The all-to-all connectivity within each cluster is a key contributor because the qubits forming the inter-cluster links act as a bridge. Through this single connecting qubit, any qubit in one cluster can indirectly influence, and be influenced by, all the qubits in the other cluster. This may necessitate an increase in the bond dimension of the virtual bond connecting the clusters for an accurate representation of the updated state. This issue is analogous to the time evolution of many-body systems where correlations spread throughout the system as time progresses. Accurately capturing these long-range correlations often requires increasing the bond dimension in the simulation.

 \begin{figure}[!t]
\centering
\includegraphics[width=1\columnwidth]{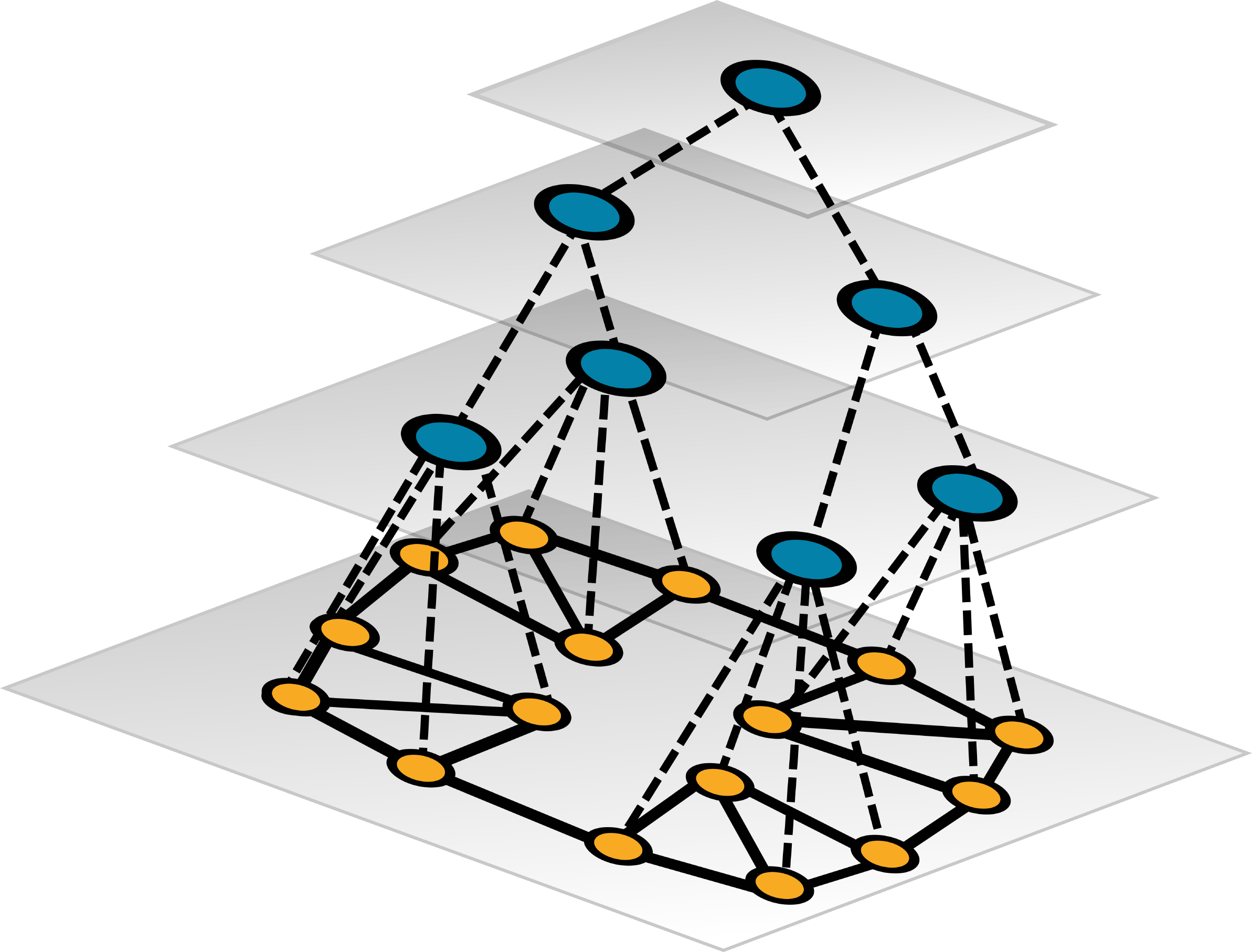}
\caption{The bridged 3-regular graph representing the two-qubit gate connectivity is embedded into a hierarchical tree structure. Clusters of qubits (orange nodes) in the lowest level make up the distinct leaf nodes of the tree tensor network. These clusters are then coupled to form the internal nodes of the upper layer while capturing the inter-cluster two-qubit gate connectivity.} 
\label{Fig: TTN with 3-reg}
\end{figure}

\subsection{QAOA}
\label{sec: Results qaoa}

To test our method on a useful task, we simulate {the quantum approximate optimization algorithm (QAOA) \cite{farhiQAOA,QO_VA,QAOA_Lukin,farhiQAOA2}. QAOA is a hybrid quantum-classical variational
algorithm designed to tackle combinatorial optimization problems. In the QAOA framework, the optimization task is mapped to a physical system by encoding the problem's cost function into a Hamiltonian whose ground state represents the optimal solution. The algorithm proceeds by preparing a parameterized quantum state and using a classical optimizer to adjust the parameters in order to minimize the energy of the Hamiltonian.}

In this section, we focus on the QAOA circuit on various graphs for the MaxCut problem. The MaxCut problem involves partitioning the vertices of a graph into two subsets to maximize the number of edges crossing between these subsets. The QAOA circuit for the MaxCut problem can on a graph with $N$ vertices with depth $p$ and $2p$ parameters \cite{farhiQAOA,IsingForm_NP,QAOA_MaxCut,QAOA_Lukin} is expressed as 
\begin{equation}
\bigl\vert\boldsymbol{\theta}\equiv\bigl\{\boldsymbol{\beta},\boldsymbol{\gamma}\bigr\}\bigr\rangle=\prod_{\ell=1}^pU_{\beta_\ell}U_{\gamma_\ell}H^{\otimes N}\vert 0\rangle^{\otimes N},
\end{equation}
where $U_{\beta_\ell}=\prod\nolimits_j\exp\bigl(-i\frac{\beta_\ell}{2}X_j\bigr)$ with $X_j$ the Pauli operator on qubit $j$, $U_{\gamma_\ell}=\exp\bigl(-i\frac{\gamma_\ell}{2}C\bigr)$ with $C=\sum_{\{i,j\}\in E}w_{ij}Z_iZ_j$  the cost function operator of the MaxCut problem running over the edges $\{i, j\}$ on the graph and $H$ the Hadamard gate applied on the individual qubits. Here, we want to probe how accurately the QAOA circuit can be simulated. Therefore, the variational parameters of the circuit are not optimized. Instead, the $2p$ parameters are sampled from uniform distributions: $\gamma\in [0,2\pi)$ and $\beta\in [0,\pi)$. We simulate the QAOA algorithm on two types of graphs: 3-regular graphs and graphs with two-qubit gate connectivity that align with the structure of tree-like circuits. For both the graphs, weights are set to be uniform as $w_{ij}=w=1$.

\subsubsection{3-regular graphs}

 A 3-regular graph with $N$ vertices is a graph where each vertex is connected to three other and distinct vertices at random without having self-loops or parallel edges. A TTN can be constructed to effectively leverage the underlying connectivity of such graphs. This approach allows obtaining high fidelity even at relatively small bond dimensions. An example of such a construction is shown in Fig.~\ref{Fig: TTN with 3-reg}, where clusters of four qubits are identified based on their edge density. These clusters then form the nodes associated with the lowest layer of the TTN. These clusters are then coupled together to form the nodes of the upper layer while capturing the inter-cluster connectivity.

\begin{figure}[!t]
\centering
\includegraphics[width=1\columnwidth]{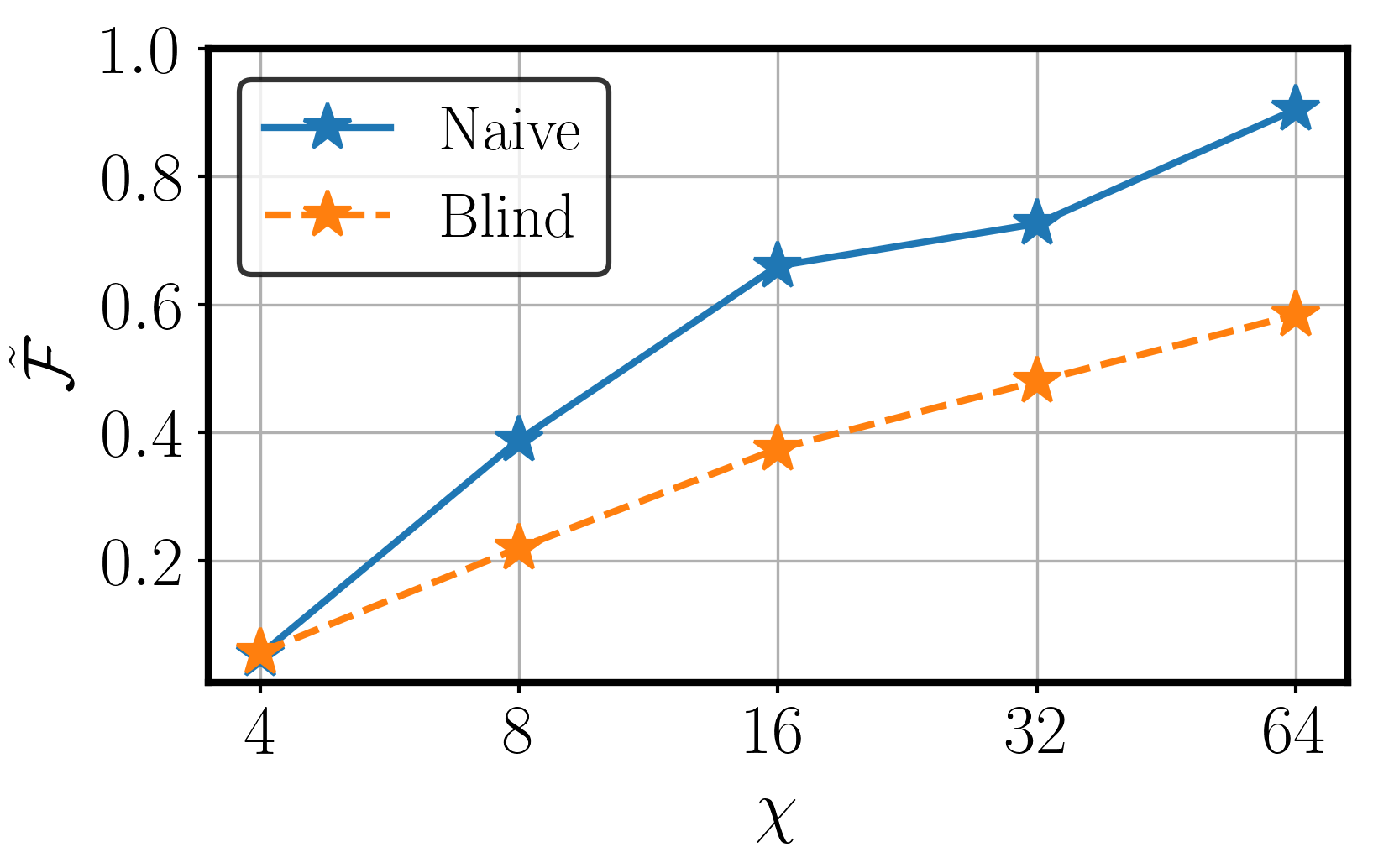}
   \caption{Comparison of fidelities $\mathcal{\tilde{F}}$ as a function of bond dimension $\chi$ obtained from the simulation of QAOA circuits on random 3-regular graphs using TTN constructed via blind and naive approaches. The results are shown for $N=64$ at depth $p=1$ with $N_{2g}=96$ two-qubit gates. } 
   \label{Fig: Naive_vs_Blind}

\end{figure}

For random 3-regular graphs, we employ a ``naive" clustering approach to construct the TTN. Clusters are first identified within the graph such that there is a high density of edges within each cluster and only a few edges connecting different clusters. This clustering is achieved using the algorithm described in \cite{Newman_community_detection}, which is designed for community detection in networks. Once the clusters are identified, qubits belonging to the same cluster are positioned closer together in the TTN. {This procedure effectively acts as a qubit reordering strategy which reduces the number of virtual bonds connecting qubits that belong to the same cluster.} However, this approach does not account for inter-cluster relationships, which limits its ability to fully exploit the graph's connectivity. We compare this naive approach to a ``blind" one that doesn't use any structural information about the graph during TTN construction, {while maintaining the same TTN architecture}. Despite its simplicity, the naive method demonstrates significant improvements in fidelity compared to the blind approach, as shown in Fig.~\ref{Fig: Naive_vs_Blind}. The results are presented for the simulation of a QAOA circuit with 64 qubits at depth $p=1$ on random 3-regular graphs. The naive approach achieves better results with \( \chi=16 \) than the blind approach with \( \chi=64 \), highlighting the effectiveness of leveraging even basic structural information in TTN construction can enhance algorithm performance. These findings suggest that further improvements in fidelity can be achieved with more sophisticated TTN constructions that exploit both intra- and inter-cluster relationships.

\begin{figure}[!t]
\centering
\includegraphics[width=1\columnwidth]{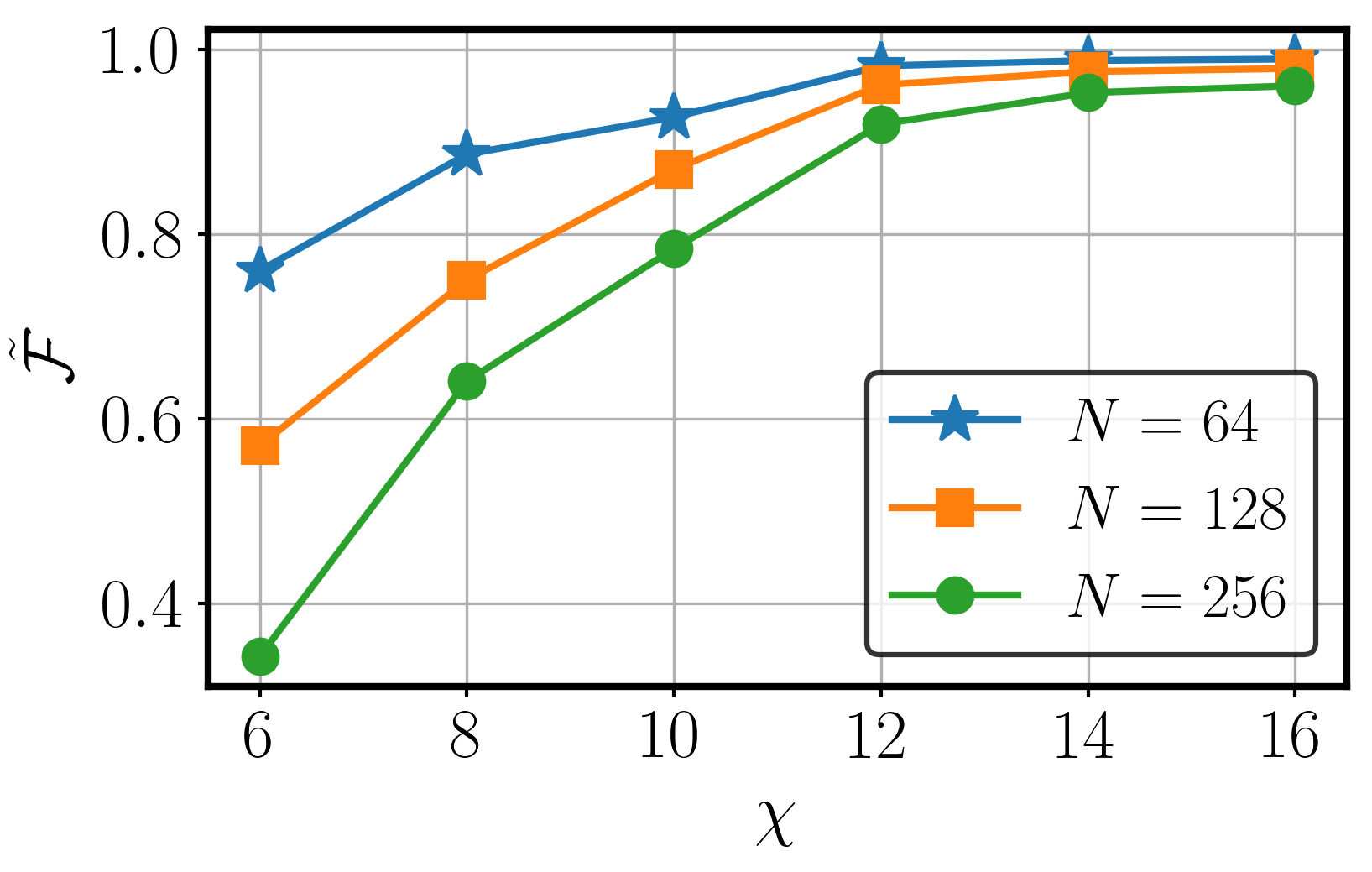}
   \caption{Fidelity $\mathcal{\tilde{F}}$ as a function of bond dimension $\chi$ obtained from the simulation of QAOA circuits on bridged 3-regular graphs at circuit depth $p=4$. The results are shown for $N_{2g}=384,768,$ and $1536$ two-qubit gates, corresponding to systems of $N=64,128,$ and $256$ qubits, respectively.  } 
   \label{Fig: Bridged 3-regular}
\end{figure}

 A subclass of 3-regular graphs that has built-in clustering is the bridged 3-regular graphs. In such graphs, nodes form tightly connected clusters that are arranged in a circular chain. This inherent clustering allows for constructing a TTN that aligns with the graph connectivity as illustrated in Fig.~\ref{Fig: TTN with 3-reg} for a bridged 3-regular graph with sixteen vertices. To evaluate the performance of our method, we perform QAOA simulations on such graphs. Fig.~\ref{Fig: Bridged 3-regular} shows results at depth $p=4$ for $N=64,128,$ and $256$ qubits. We achieve fidelities of $\mathcal{\tilde{F}}\approx1$ with very small bond dimensions across all system sizes. These results demonstrate the efficiency of TTNs when applied to graphs that admit hierarchical embedding into a tree network.  

\subsubsection{Tree-like circuits}
\begin{figure}[!t]
    \centering
    \includegraphics[width=1\columnwidth]{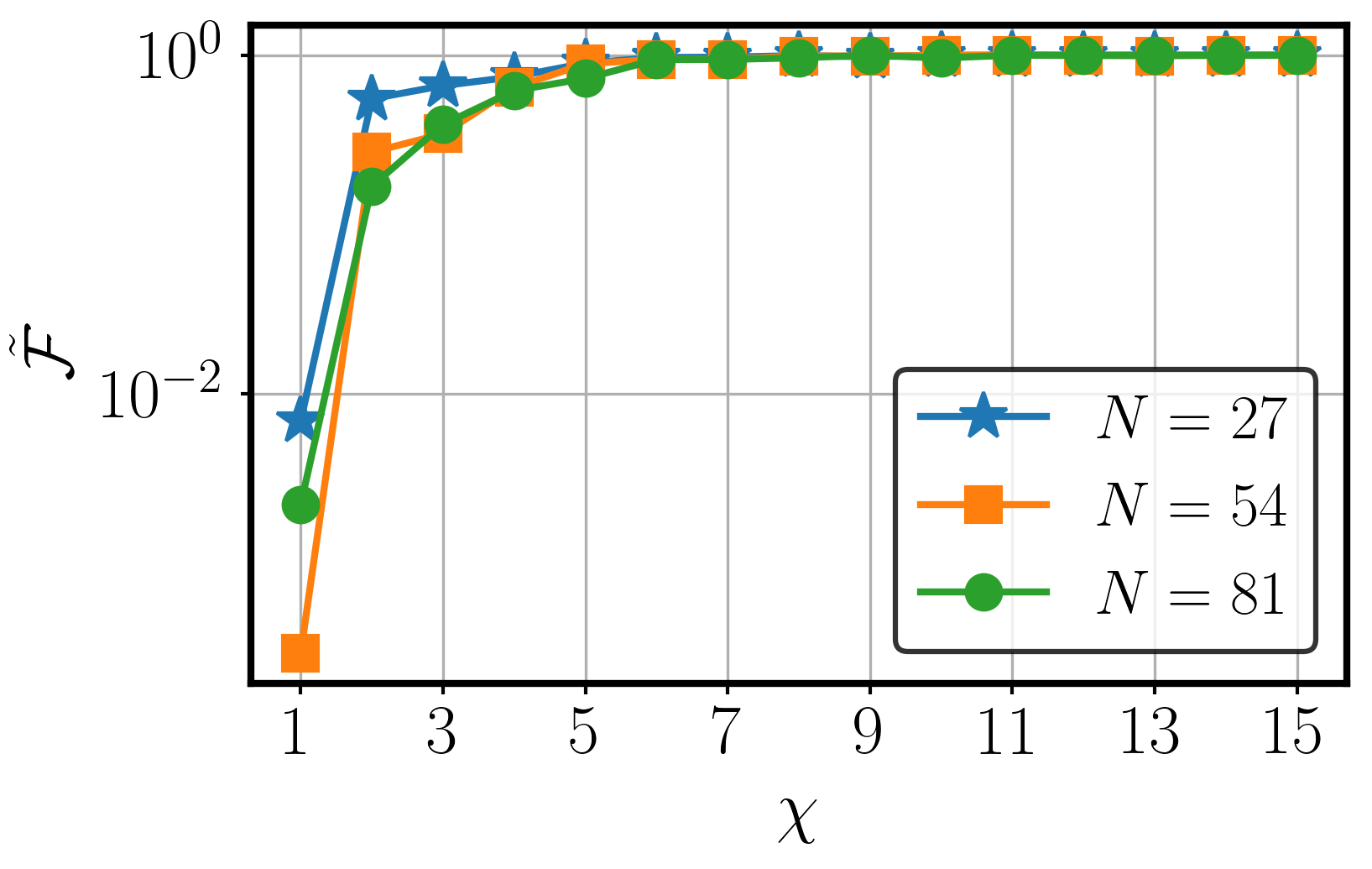}
    \caption{Fidelity $\mathcal{\tilde{F}}$ as a function of bond dimension $\chi$ obtained from the simulation of QAOA tree-like circuits at circuit depth $p=4$. The results are shown for $N_{2g}=156,316,$ and $480$ two-qubit gates, corresponding to systems of $N=27,54,$ and $81$ qubits, respectively.}
    \label{Fig: QAOA tree-like}
\end{figure}

We now focus on graphs whose edges correspond to the two-qubit gate connectivity of tree-like circuits. As previously discussed, these graphs exhibit hierarchical tree-like clustering. This hierarchical structure makes TTN particularly well-suited for simulating QAOA circuits aimed at solving combinatorial problems on such graphs. As shown in Fig.~\ref{Fig: QAOA tree-like}, we achieve fidelities of $\mathcal{\tilde{F}}\approx1$ at a very low bond dimension of $\chi=8$ across all system sizes considered for a QAOA circuit with depth $p=4$.

\begin{figure}[!t]
    \centering
    \includegraphics[width=1\columnwidth]{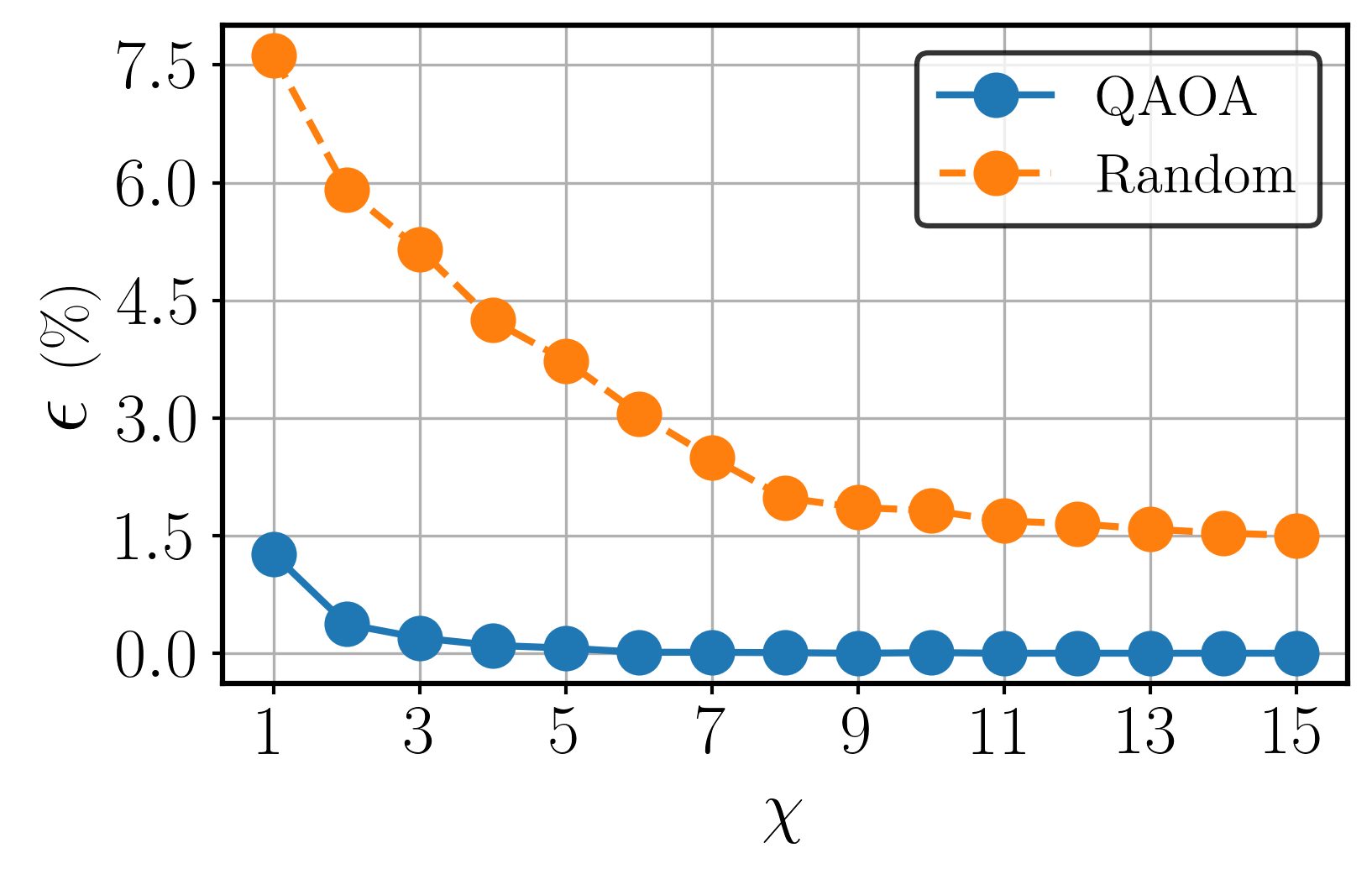}
    		\caption{Comparison of error rates $\epsilon$ as a function bond dimension $\chi$ obtained from the simulation of random and QAOA tree-like circuits with $N_{2g}=480$ two-qubit gates for a system of $N=81$ qubits.} 
    \label{Fig: QAOAvsRandom}
\end{figure}

Quantum circuits designed for practical applications often utilize a particular type of parametrized two-qubit gate throughout the circuit. For example, QAOA circuits employ carefully chosen two-qubit gates tailored for a specific problem. This contrasts with random circuits that use two-qubit gates that are completely random unitaries and different from each other. For instance, in the QAOA circuit for the MaxCut problem, the same type of two-qubit gate is applied consistently at each depth $p$, varying only by a parameterized angle. This inherent structure in the gates results in significantly less entanglement generation compared to random circuits. Consequently, this allows for a more accurate representation of the quantum state using the DMRG algorithm. To highlight this contrast, we compare the error per gate from simulations of tree-like random and tree-like QAOA circuits with $N_{2g} = 480$ two-qubit gates for a system of $N = 81$ qubits. As shown in Fig.~\ref{Fig: QAOAvsRandom}, the QAOA simulations achieve significantly lower error rates. Specifically, the error is reduced from $\epsilon\approx1.5\%$ for the random circuit to $\epsilon\approx0.0004\%$ for the QAOA circuit with $\chi=15$, corresponding to a reduction by a factor of nearly 4000.

\section{Conclusion and Outlook}\label{Sec: Conclusion}

Tensor networks offer {an} ideal framework for classical simulation of quantum circuits and have been successfully applied in various contexts \cite{TEBD_Miles,Gaussian_boson_sampling_TN,TTN_Mendl,Grover_Miles,PEPS_random_circuits,FFT_Miles,QC_simulation_Batch_TN,Miles_kicked_ising}. In particular, the DMRG algorithm was generalized to simulate quantum circuits with a finite fidelity using matrix product states \cite{DMRG_Miles}. However, despite their efficiency, MPS are fundamentally limited by their chain-like structure. This structural constraint restricts their ability to represent quantum states with more complex entanglement patterns. To address this limitation, we extend the DMRG algorithm for quantum circuit simulation to tree tensor networks.

In this study, we considered the simulation of random circuits and QAOA circuits across various two-qubit gate connectivities, including a special class of circuits referred to as \textit{tree-like} circuits. The key implication of our work is that {the DMRG aglorithm with} TTNs provides a promising framework for simulating quantum circuits, especially when gate connectivities exhibit some form of clustering or hierarchical structure. For the tree-like circuits with random gates, we showed that MPS requires substantially more memory than TTN to achieve comparable error per gate. This advantage is due to TTN's ability to better capture long-range correlations than MPS. For QAOA circuits on random 3-regular graphs, we demonstrated that a naive TTN construction approach based on clustering graph nodes greatly improves fidelity compared to a blind approach that ignores structural information. To further emphasize the 
importance of exploiting structure, we simulated the QAOA circuit for bridged 3-regular graphs and graphs with two-qubit gate connectivity that aligns with the structure of tree-like circuits. For both types of graphs, we achieved high fidelities at very small bond dimensions. Our results also highlight that circuits designed for practical tasks, such as QAOA, are significantly more amenable to compression than random circuits.

Our findings demonstrate that the {DMRG algorithm with} tree tensor networks offers an efficient and flexible framework for simulating quantum circuits. A natural extension of our work is to consider a two-site version of the compression step. Although the previous work using MPS reported no significant improvement from the two-site variant over the single-site version \cite{DMRG_Miles}, its potential advantages in the TTN setting remain worth investigating. {In this work, we employed layer-wise regular tree structures. Future research could explore and compare more general tree topologies by leveraging more advanced TTN constructions \cite{TTN_Mendl,TTNOPT_GS,TTN_QC_opt,TTN_QC_opt2}}. Further performance gains could be achieved by enabling adaptive changes to the TTN structure during simulation, as developed for MPS in \cite{Cluster_DMRG}. Notably, the compression step in our algorithm is quite general and can be incorporated into other TTN-based quantum circuit algorithms, such as those proposed in Ref.~\cite{TTN_Mendl}.  Recent works, such as Quantinuum's realization of random quantum circuits in arbitrary geometries and benchmarking them with MPS \cite{Random_circ_Quantinuum}, highlight the relevance of our approach as an alternative benchmarking tool. The framework employed here can also be extended to simulate the time evolution of a quantum many-body system \cite{In_prep}. Finally, for certain applications like QAOA circuits, where the entanglement generated remains relatively low \cite{Entanglement_QAOA}, our method offers a particularly efficient and practical solution.

\begin{acknowledgments}
Z.Z. and P.S. are funded by the Cluster of Excellence
``CUI: Advanced Imaging of Matter'' of the Deutsche Forschungsgemeinschaft (DFG) - EXC 2056 - Project ID 390715994. The tensor contractions were performed using the \texttt{cotengra} \cite{Cotengra} and \texttt{opt\_einsum} \cite{
opt_einsum} libraries. The qubit connectivity graphs were generated using the \texttt{networkX} \cite{networkx} library.
\end{acknowledgments}

\appendix
\section{Comparison with SVD-based approach}
\label{Appendix: SVD compare}
{
In this section, we compare the DMRG method used in our work with the SVD-based truncation scheme described in Ref.~\cite{TTN_Mendl}. The SVD-based approach approximates the quantum state by applying an SVD after each gate operation and discarding singular values below a certain threshold. The truncations introduce an error with each gate. These errors compound with subsequent gate applications, potentially leading to a significant loss of fidelity.}

{In contrast, the DMRG method employs a variational compression scheme. It iteratively optimizes each tensor to find the best possible TTN representation of the evolved quantum state for a given bond dimension. This variational approach is superior to the SVD-based compression method as it minimizes the distance to the target state and avoids the compounding errors associated with successive SVD truncations \cite{SCHOLLWOCK_TN_bible,DMRG_Miles}. }
\begin{figure}[!t]
\centering
\includegraphics[width=\columnwidth]{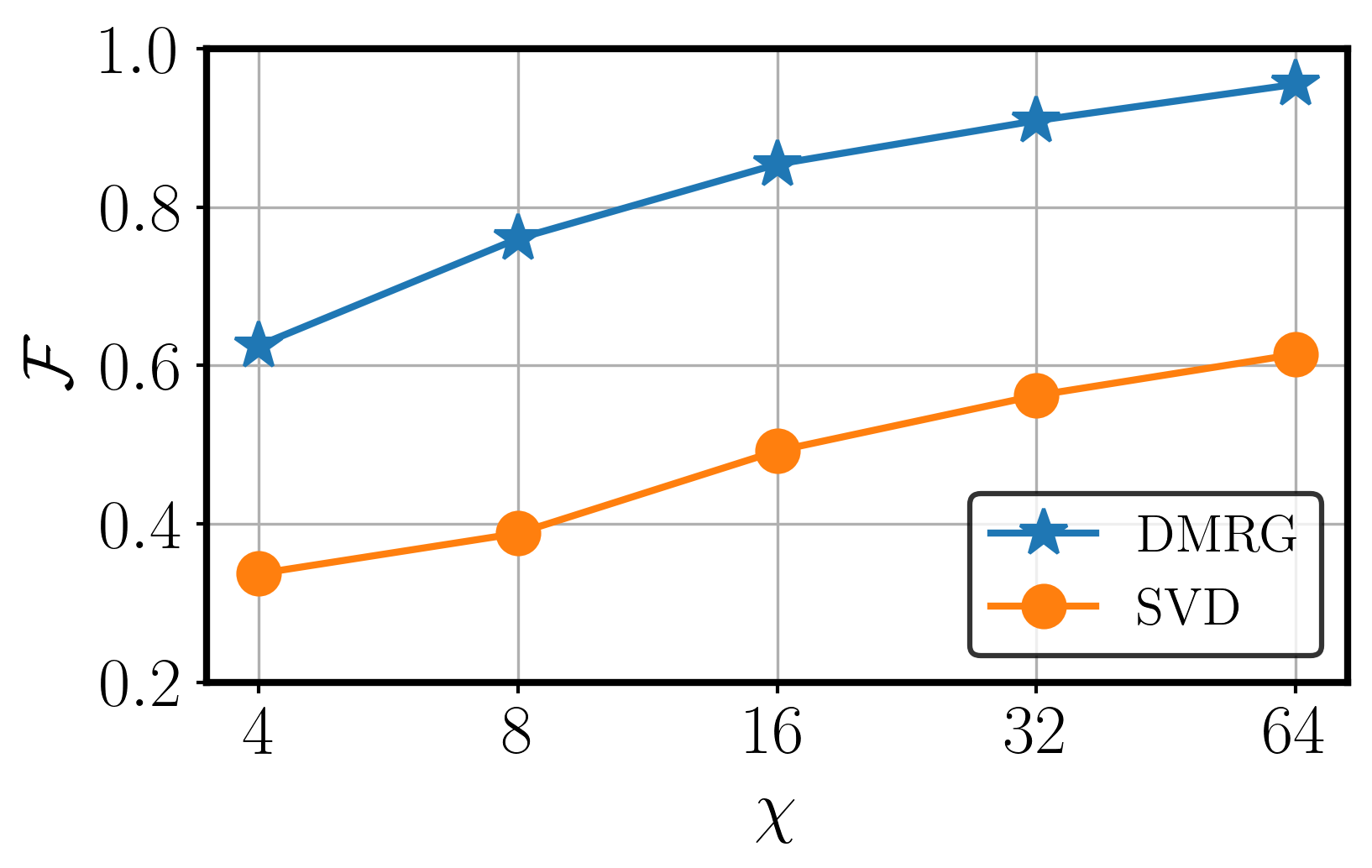}
\caption{Comparison of fidelities $\mathcal{F}$ as a function of bond dimension $\chi$ obtained from the simulation of QAOA circuits on random 3-regular graphs by the DMRG method and SVD-TTN from Ref. \cite{TTN_Mendl}. A binary tree structure is employed for representing the quantum state in both cases. No information regarding the graph is exploited. Results are shown for $N = 16$ at depth $p = 1$ with $N_{2g} = 24$ two-qubit gates.}
\label{Fig:Blind}
\end{figure}

\begin{figure}[!t]
\centering
\includegraphics[width=\columnwidth]{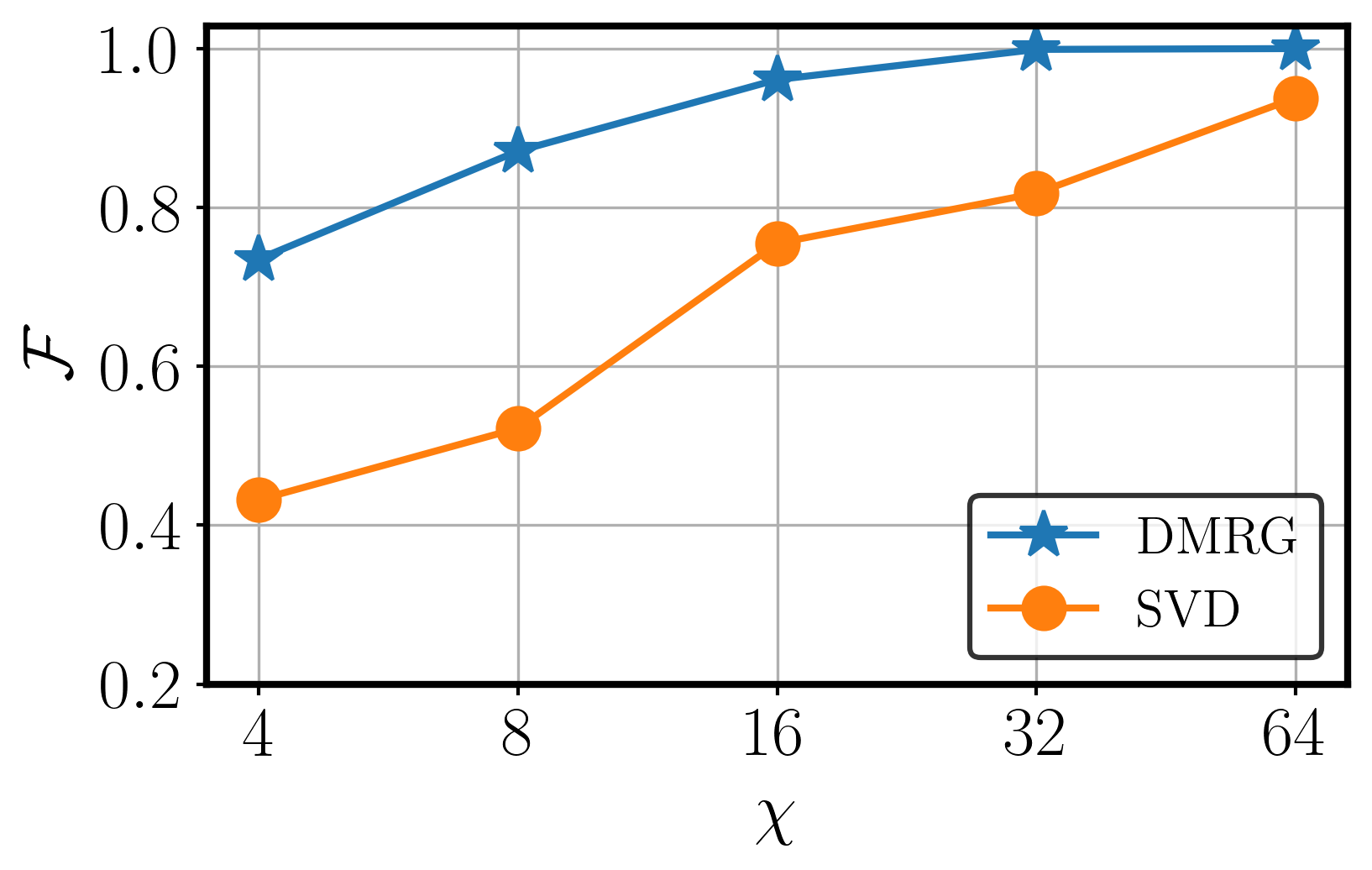}
\caption{Comparison of fidelities $\mathcal{F}$ as a function of bond dimension $\chi$ obtained from the simulation of QAOA circuits on random 3-regular graphs by the DMRG method and SVD-TTN from Ref. \cite{TTN_Mendl}. A binary tree structure is used for both methods. The naive method described in Sec.~\ref{sec: Results qaoa} is applied for qubit reordering to improve ansatz expressivity. Results are shown for $N = 16$ at depth $p = 1$ with $N_{2g} = 24$ two-qubit gates.}
\label{Fig:Naive}
\end{figure}
{To illustrate these differences quantitatively, we simulate quantum approximate optimization algorithm (QAOA) circuits for the Max-Cut problem on random 3-regular graphs with $N=16$ qubits using both methods.  The fidelity $\mathcal{F}$ of each method is computed relative to exact statevector simulations performed using  \texttt{Qiskit} \cite{qiskit}. The choice of $N=16$ enables this exact benchmarking. To ensure a fair comparison, both methods are implemented using the same binary TTN, and fidelities are reported as a function of the bond dimension. Results for Ref. \cite{TTN_Mendl} were obtained using the publicly available code provided by the authors.}

\begin{figure}[!t]
    \centering
    \includegraphics[width=1\columnwidth]{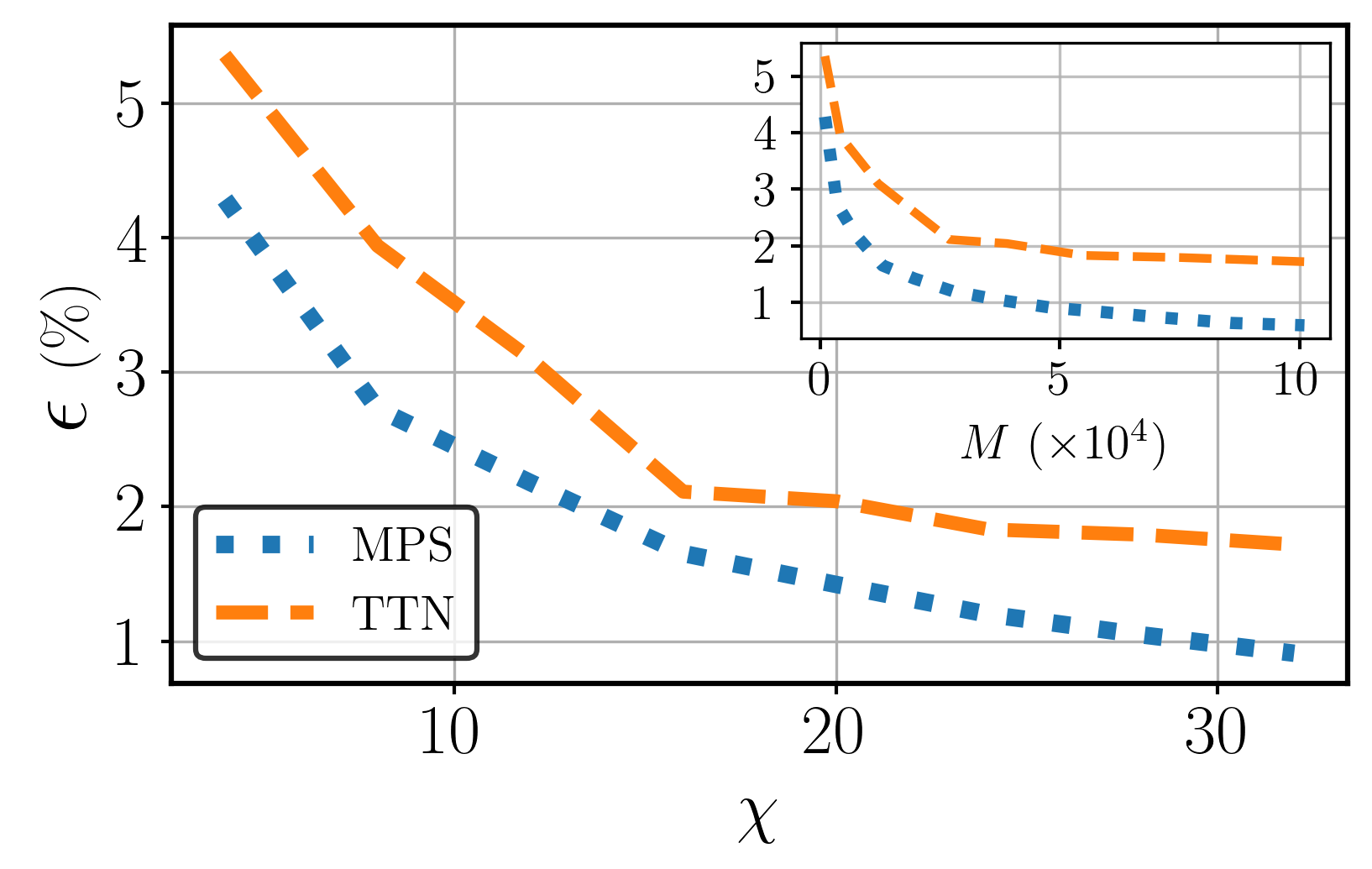}
    \caption{Comparison of error rates $\epsilon$ as a function of bond dimension $\chi$ obtained from the simulation of random nearest-neighbor circuits using MPS and TTN. The results are for a system of $N=32$ qubits with $N_{2g}=310$ two-qubit gates. The inset shows error rates $\epsilon$ as a function of memory footprint $M$ obtained from the same simulation.} 
    \label{Fig: Error NN Compare}
\end{figure}

{Fig.~\ref{Fig:Blind} presents a “blind” comparison, where both methods use a binary TTN ansatz without exploiting any information about the graph structure. Even at a very low bond dimension of $\chi = 4$, our DMRG-based method outperforms the SVD-based approach at $\chi = 64$, demonstrating the superior accuracy of variational optimization.}

\begin{figure}[!h]
    \centering
    \includegraphics[width=1\columnwidth]{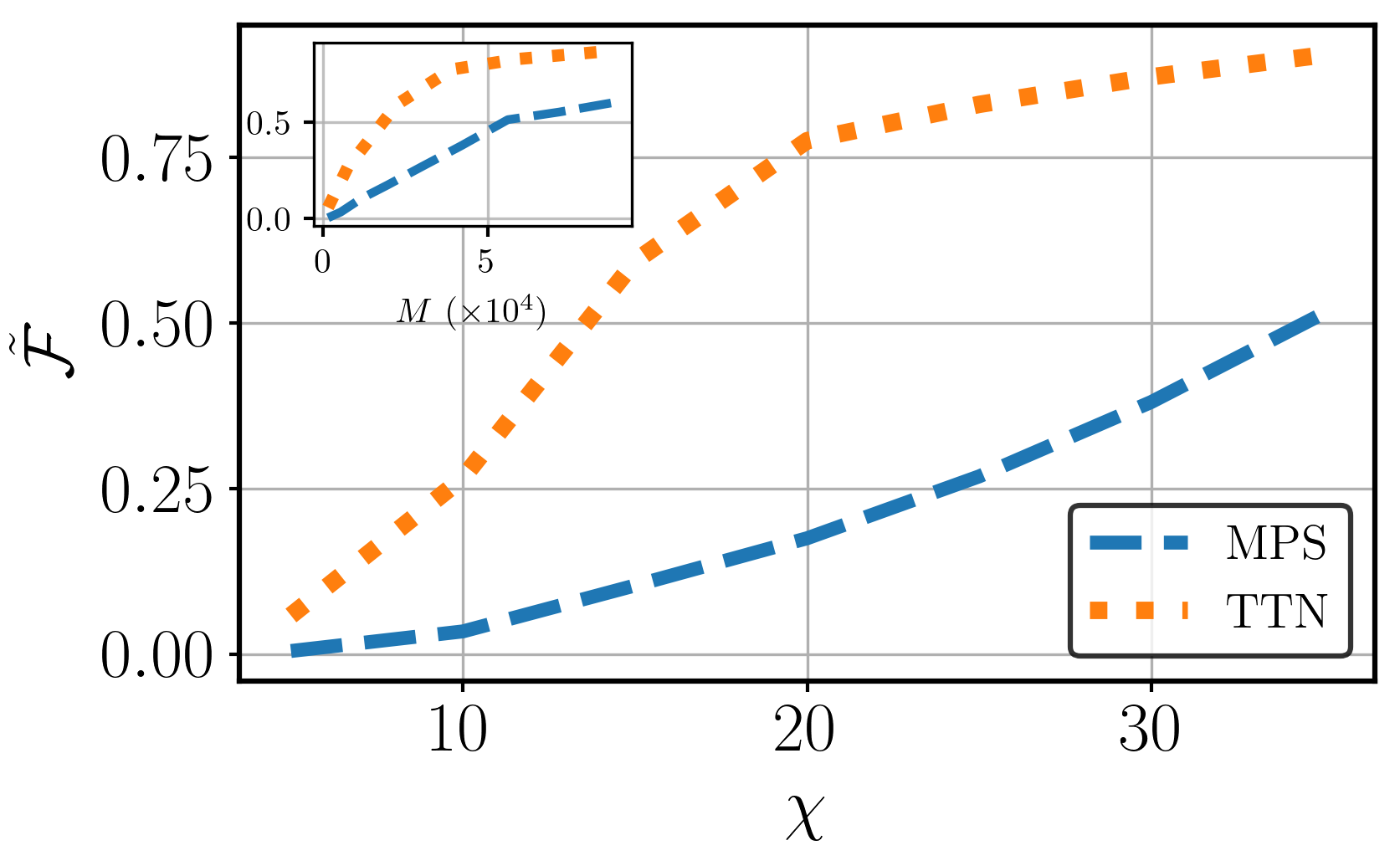}
    \caption{Comparison of fidelities $\mathcal{\tilde{F}}$ as a function of bond dimension $\chi$ obtained from the simulation of tree-like QAOA circuits using MPS and TTN. The results are for a system of $N=32$ qubits at depth $p=15$ with $N_{2g}=825$ two-qubit gates. {The inset shows fidelity $\mathcal{\tilde{F}}$ as a function of memory footprint $M$ obtained from the same simulation.} } 
    \label{Fig: Treelike Compare}
\end{figure}

{Fig.~\ref{Fig:Naive} shows the same comparison, now incorporating a “naive” qubit reordering strategy as Sec.~\ref{sec: Results qaoa}. In both cases, a binary TTN is used, but the strongly correlated qubits are placed together in the TTN. While both approaches benefit from reordering, our DMRG-based algorithm at $\chi = 16$ still outperforms Ref. \cite{TTN_Mendl} at $\chi = 64$. The relative improvement is more pronounced in Ref.\cite{TTN_Mendl}, indicating that the SVD-based approach is more sensitive to the choice of TTN structure and less robust to long-range entangling gates. This comparison highlights the advantage of our variational framework, which achieves higher accuracy at lower bond dimension while maintaining greater resilience to circuit nonlocality.}

\section{Comparison between MPS and TTN}
\label{Appendix: Ansatz compare}
{We compare MPS and TTN as variational ansatz within the DMRG algorithm for simulating the quantum circuits. All simulations are performed on a system of $N = 32$ qubits. For TTN, we adopt a binary tree structure. It is important to note that a direct comparison between TTN and MPS is inherently challenging. As a result, our analysis primarily focuses on the bond dimension $\chi$, which determines the maximum half-chain entanglement entropy that each ansatz can represent \cite{SCHOLLWOCK_TN_bible}. We also provide a comparison of the memory footprint.}

{For random circuits, we consider nearest-neighbor connectivity among two-qubit gates.  The linear connectivity of these circuits aligns naturally with the structure of MPS which makes them ideal for such simulations. In contrast, TTNs impose a hierarchical grouping of qubits that introduces a structural bias that is not compatible with the geometry of these circuits. This mismatch is evident in Fig.~\ref{Fig: Error NN Compare}, where MPS outperforms TTNs. The performance advantage observed for MPS in linear circuits is similar to the advantage TTNs exhibit in tree-like circuit topologies (see Fig.~\ref{sec: Results random}). These results highlight the importance of choosing an ansatz compatible with the circuit structure for efficient simulation.}

{For QAOA circuits, we consider both tree-like circuits and random 3-regular graphs. The hierarchical structure of tree-like circuits makes them naturally compatible with the TTN ansatz. As shown in Fig.~\ref{Fig: Treelike Compare}, TTN significantly outperforms MPS for these circuits. This outcome is consistent with expectations, as the linear geometry of MPS limits its ability to efficiently capture the entanglement present in hierarchically structured circuits.}

\begin{figure}[!t]
    \centering
    \includegraphics[width=1\columnwidth]{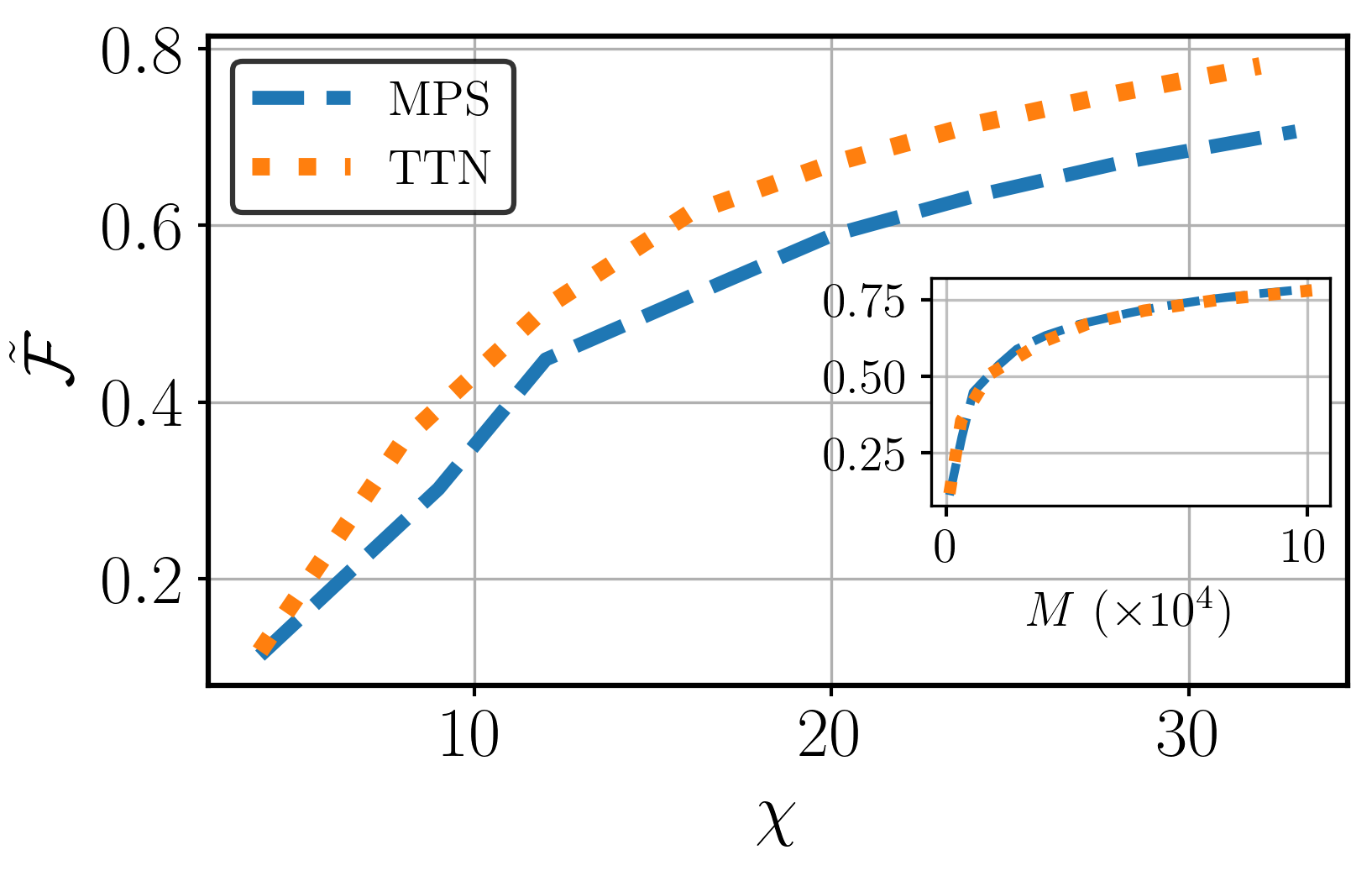}
    \caption{Comparison of fidelities $\mathcal{\tilde{F}}$ as a function of bond dimension $\chi$ obtained from the simulation of QAOA circuits on random 3-regular graphs using MPS and TTN. The results are for a system of $N=32$ qubits at depth $p=2$ with $N_{2g}=96$ two-qubit gates. The inset shows fidelity $\mathcal{\tilde{F}}$ as a function of memory footprint $M$ obtained from the same simulation.} 
    \label{Fig: 3-reg Compare}
\end{figure}
{For QAOA circuits on random 3-regular graphs, we apply a naive qubit reordering strategy, as described in Sec.\ref{sec: Results qaoa}, to both MPS and TTN simulations. As shown in Fig.\ref{Fig: 3-reg Compare}, TTN slightly outperforms MPS in terms of bond dimensions. However, the inset reveals that the overall memory footprints of the two ansatzes are comparable. This similarity may result from the limitations of the naive reordering scheme, which does not fully exploit the structure of the graph. More advanced TTN construction methods could better capture the graph topology and provide a more efficient representation. In contrast, for MPS, performance improvements are limited unless the graph exhibits linear clustering. A more comprehensive study is warranted to fully assess the trade-offs between the two approaches in general graph-based QAOA instances.}

\bibliography{REF} 
\end{document}